\documentclass{article}
\usepackage[T1]{fontenc}
\usepackage{microtype}
\usepackage{graphicx}
\usepackage{booktabs}
\usepackage{hyperref}

\PassOptionsToPackage{noend}{algorithmic}
\IfFileExists{latexml.sty}{\usepackage{latexml}}{\newif\iflatexml}
\iflatexml
\edef\ltxstyle{icml}
\expandafter\usepackage\expandafter{\ltxstyle}
\newenvironment{mlsysauthorlist}{}{}
\let\mlsystitle\icmltitle
\let\mlsystitlerunning\icmltitlerunning
\let\mlsysauthor\icmlauthor
\let\mlsysaffiliation\icmlaffiliation
\let\mlsyscorrespondingauthor\icmlcorrespondingauthor
\let\mlsyskeywords\icmlkeywords
\else
\usepackage[accepted]{mlsys2025}
\makeatletter\renewcommand{\Notice@String}{Preprint. Under review.}\makeatother
\fi
\AtBeginDocument{\hypersetup{pdfsubject={}}}

\usepackage{amsmath,amssymb}
\usepackage{array,calc}
\usepackage{adjustbox}
\usepackage{textcomp}
\usepackage{newunicodechar}
\usepackage{tikz}
\usetikzlibrary{arrows.meta, calc, decorations.pathreplacing}

\definecolor{figink}{HTML}{1B1B1F}
\definecolor{figgrid}{HTML}{D5D5DC}
\definecolor{figmuted}{HTML}{5A5A66}
\definecolor{figgood}{HTML}{1F6F4A}
\definecolor{figbad}{HTML}{9B2226}
\definecolor{fignil}{HTML}{8A8A95}
\definecolor{figgpu}{HTML}{2B5FA8}
\definecolor{figcpu}{HTML}{B8641A}
\definecolor{figfaint}{HTML}{F4F4F7}
\definecolor{figlaunch}{HTML}{3A3A44}
\definecolor{figwait}{HTML}{E9E9EE}
\definecolor{figslate}{HTML}{3D4A5C}
\definecolor{figpurple}{HTML}{8A4F9E}
\definecolor{figtint}{HTML}{E8F0FA}
\definecolor{figcputint}{HTML}{FBEEE2}
\definecolor{figpad}{HTML}{F1F1F4}

\newcommand{\figtitle}{\fontsize{8}{9.5}\selectfont\sffamily\bfseries}
\newcommand{\figlabel}{\fontsize{7.5}{9}\selectfont\sffamily}
\newcommand{\figtext}{\fontsize{7}{8.5}\selectfont\sffamily}
\newcommand{\figsmall}{\fontsize{6.5}{7.5}\selectfont\sffamily}

\makeatletter
\newcommand{\FigHatch}{%
  \pgfpointanchor{path picture bounding box}{south west}%
  \edef\fig@xa{\strip@pt\pgf@x}\edef\fig@ya{\strip@pt\pgf@y}%
  \pgfpointanchor{path picture bounding box}{north east}%
  \edef\fig@xb{\strip@pt\pgf@x}\edef\fig@yb{\strip@pt\pgf@y}%
  \pgfmathtruncatemacro\fig@ma{floor((\fig@xa-\fig@yb)/3.1113)}%
  \pgfmathtruncatemacro\fig@mb{ceil((\fig@xb-\fig@ya)/3.1113)}%
  \foreach \fig@m in {\fig@ma,...,\fig@mb}
    {\draw[fignil, line width=0.35pt] ({\fig@m*3.1113pt+\fig@ya pt}, \fig@ya pt) -- ({\fig@m*3.1113pt+\fig@yb pt}, \fig@yb pt);}}
\makeatother

\tikzset{
  fig/.style={font=\figtext, text=figink, line cap=butt},
  seg/.style={draw=#1, fill=#1, line width=0.4pt},
  waitfill/.style={fill=figwait, draw=fignil, line width=0.4pt},
  hatched/.style={waitfill, path picture={\FigHatch}},
  lane/.style={draw=figgrid, line width=0.5pt},
  stepend/.style={draw=figink, line width=0.6pt, dash pattern=on 1.5pt off 1.5pt},
  readback/.style={figbad, line width=0.7pt, -{Triangle[length=3.2pt, width=3.2pt]}},
  copyto/.style={figink, line width=0.6pt, -{Triangle[length=3.2pt, width=3.2pt]}},
  flow/.style={figink, line width=0.6pt, -{Triangle[length=3.6pt, width=3.6pt]}},
}

\newcommand{\FigBox}[6][]{%
  \path[draw=figink, fill=white, line width=0.6pt, rounded corners=1.6pt, #1]
    ({#2}, {#3}) rectangle ++({#4}, {#5});
  \node[inner sep=0, align=center] at ({#2+(#4)/2}, {#3+(#5)/2}) {#6};}
\newcommand{\FigBar}[6][]{%
  \path[line width=0.4pt, #1] ({#2}, {#4-(#5)/2}) rectangle ({#3}, {#4+(#5)/2});
  \node[inner sep=0, font=\figsmall] at ({(#2+#3)/2}, {#4}) {#6};}
\newcommand{\FigCell}[6][]{%
  \path[draw=figgrid, fill=white, line width=0.6pt, #1] ({#2}, {#3-(#5)/2}) rectangle ++({#4}, {#5});
  \node[inner sep=0, font=\fontsize{7}{8}\selectfont\ttfamily] at ({#2+(#4)/2}, {#3}) {#6};}
\newcommand{\FigNote}[4][]{%
  \node[anchor=west, inner sep=0, align=left, text=figmuted, #1] at ({#2}, {#3}) {#4};}
\tikzset{
  tintbox/.style={draw=figgpu, fill=figtint, line width=0.9pt},
  cputintbox/.style={draw=figcpu, fill=figcputint, line width=0.9pt},
  gpubox/.style={draw=figgpu, line width=0.6pt},
  cpubox/.style={draw=figcpu, line width=0.6pt},
  padcell/.style={fill=figpad, path picture={\FigHatch}},
  thickflow/.style={figink, line width=0.8pt, -{Triangle[length=4pt, width=4pt]}},
}

\newcommand{\ValReplayLight}{53.9}
\newcommand{\ValReplaySaturated}{91.1}
\newcommand{\ValReplayTwentySlots}{57.3}
\newcommand{\ValRaggedGpu}{+113.4}
\newcommand{\ValPaddedCpu}{+4.2}
\newcommand{\ValRaggedCpu}{$-$9.7}
\newcommand{\ValMedianPageTokens}{843}
\newcommand{\ValPaddedWalkSlowdown}{22}
\newcommand{\ValPaddedSlots}{8{,}192}

\providecommand{\tightlist}{\setlength{\itemsep}{0pt}\setlength{\parskip}{0pt}}

\newcommand{\thead}[1]{\textbf{#1}}
\renewcommand{\algorithmiccomment}[1]{\hfill\(\triangleright\) #1}

\newunicodechar{−}{\textminus}
\newunicodechar{×}{\texttimes}
\newunicodechar{¹}{\textsuperscript{1}}
\newunicodechar{⁶}{\textsuperscript{6}}
\newunicodechar{⁷}{\textsuperscript{7}}
\newunicodechar{⁸}{\textsuperscript{8}}
\newunicodechar{⁹}{\textsuperscript{9}}
\newunicodechar{⁰}{\textsuperscript{0}}

\mlsystitlerunning{LEGIT: Constrained Decoding of Heterogeneous Pages}

\begin{document}

\twocolumn[
\mlsystitle{LEGIT: Live Eligibility Gating over Identifier Tries for
Constrained Decoding of Heterogeneous Pages}

\begin{mlsysauthorlist}
\mlsysauthor{Felix Fang}{uber}
\mlsysauthor{Zheng Liu}{uber}
\mlsysauthor{Robert Saliba}{uber}
\end{mlsysauthorlist}
\mlsysaffiliation{uber}{Uber Technologies, Inc}
\mlsyscorrespondingauthor{Felix Fang}{\mbox{felixfang@uber.com}}
\mlsyscorrespondingauthor{Zheng Liu}{\mbox{zheng.liu@uber.com}}
\mlsyscorrespondingauthor{Robert Saliba}{\mbox{rsaliba@uber.com}}
\mlsyskeywords{constrained decoding, generative recommendation, tries,
LLM serving}

\vskip 0.3in

\begin{abstract}
A generative recommender names each candidate with a short token
sequence, a Semantic ID, and recommends by generating those tokens. Some
generative recommenders, including the Uber Eats search engine we are
building, write a whole results page, stores with their menu items, in
one decoding pass. Every Semantic ID on such a page must name a real
store or item that is eligible for the search, such as a store that
delivers to the user's location, and that is not already on the page.
Each item must also belong to the store above it. A trie over the
catalog keeps identifiers real, but eligibility and repetition change
with every request and token, and the grammar-constrained decoder we
tried could not enforce these changing rules exactly and quickly. We
call this constraint LEGIT, for live eligibility gating over identifier
tries. Numbering trie leaves depth first turns each rule into a range
lookup over one store trie and one item trie, and every generated page
passes an independent catalog check. On one NVIDIA B200 GPU with one or
two requests per decode step, a direct implementation that walks the
trie on the GPU makes each token about 5 times as slow as unconstrained
decoding. Across 36 implementations, hardware counters and ablations are
consistent with kernel launches and CPU-GPU waits as the source of this
slowdown. Our fastest implementation walks the trie on the CPU while the
GPU runs the model, needs no custom kernel, and cuts time per output
token by 65.9 percent against the direct GPU implementation, to about
1.7 times the unconstrained cost. This implementation is also the
fastest on a page layout that puts each item before its store.
\end{abstract}
]

\printAffiliationsAndNotice{}

\section[1. Introduction]{\hypertarget{sec:1}{}1. Introduction}\label{introduction}

A generative recommender names each candidate to recommend (a product, a
video, a menu item) with a short sequence of tokens, a Semantic ID, and
recommends by generating those tokens \citep{rajput2023tiger}. The model
is usually a decoder-only transformer, as in a large language model
(LLM), and recent industrial systems start from a pretrained LLM with
Semantic IDs added as new tokens
\citep{onerec2026onereason, he2026plum, damico2026glide}. Some of these
systems go further and let a single decoding pass write a whole result
page, so that retrieval, ranking and layout become positions in one
sequence \citep{wang2026genpage, pang2025higr, hu2026once} (\hyperlink{fig:1}{Figure 1}).
This paper studies such a system.

\begin{figure}[!tb]
\centering
\iflatexml\includegraphics[width=236.47pt]{figures/pipeline_col.svg}\else
\begin{tikzpicture}[fig, x=1in, y=1in, font=\figsmall]
\def\WA{0.02}\def\WB{3.24}\def\G{0.24}
\pgfmathsetmacro\BW{(\WB-\WA-4*\G)/5}
\newcommand{\coltitle}{\fontsize{7.4}{8.5}\selectfont\sffamily\bfseries}
\newcommand{\tok}{\fontsize{5.6}{6.5}\selectfont\ttfamily}
\tikzset{small flow/.style={figink, line width=0.6pt, -{Triangle[length=3pt, width=3pt]}}}

\def\AY{2.30}\def\AH{0.40}
\node[anchor=west, inner sep=0, font=\coltitle] at (\WA, \AY+\AH+0.15) {(a)\enspace A ranking pipeline: rules applied after ranking};
\newcommand{\stage}[3][]{%
  \FigBox[#1]{\WA+#2*(\BW+\G)}{\AY}{\BW}{\AH}{#3}
  \ifnum#2<4 \draw[small flow] ({\WA+#2*(\BW+\G)+\BW}, \AY+\AH/2) -- ({\WA+(#2+1)*(\BW+\G)}, \AY+\AH/2);\fi}
\stage{0}{request}
\stage{1}{retrieve}
\stage{2}{rank}
\stage[draw=figbad, line width=1pt]{3}{\color{figbad}apply\\\color{figbad}rules}
\stage[fill=figfaint]{4}{\bfseries the page}

\def\BY{0.04}\def\BH{1.62}
\pgfmathsetmacro\DXa{\WA+\BW+\G}
\def\PX{2.68}
\pgfmathsetmacro\DXb{\PX-\G}
\node[anchor=west, inner sep=0, font=\coltitle] at (\WA, \BY+\BH+0.15) {(b)\enspace Generative recommender: one decode};
\FigBox{\WA}{\BY+\BH/2-0.20}{\BW}{0.40}{request}
\draw[figink, line width=0.9pt] (\DXa, \BY) rectangle (\DXb, \BY+\BH);
\draw[small flow] (\WA+\BW, \BY+\BH/2) -- (\DXa, \BY+\BH/2);
\node[anchor=west, inner sep=0, font=\figsmall\itshape, text=figmuted] at (\DXa+0.07, \BY+0.13) {one model, one decode, token by token};
\FigBox[draw=figbad, line width=1pt]{\DXa+0.07}{\BY+\BH-0.50}{\DXb-\DXa-0.14}{0.40}{\color{figbad}rules checked at every token:\\\color{figbad}real, eligible, not yet on the page}
\begin{scope}[every node/.style={inner sep=0, font=\tok, anchor=west}]
  \pgfmathsetmacro\ty{\BY+\BH-0.72}
  \node[text=figmuted] (p) at (\DXa+0.08, \ty) {<page>};
  \node[text=figmuted] (s) at ([xshift=0.04in]p.east) {<store>};
  \node[text=figgpu] (sid) at ([xshift=0.04in]s.east) {s1 s2 s3 s4 s5};
  \node[anchor=north, font=\fontsize{6.2}{7}\selectfont\sffamily\itshape, text=figgpu] at ([yshift=-0.07in]sid.south) {store Semantic ID};
  \pgfmathsetmacro\ty{\ty-0.40}
  \node[text=figmuted] (i) at (\DXa+0.08, \ty) {<item>};
  \node[text=figgood] (iid) at ([xshift=0.04in]i.east) {i1 i2 i3 i4 i5};
  \node[text=figmuted] (ie) at ([xshift=0.04in]iid.east) {</item>};
  \node[text=figmuted] at ([xshift=0.04in]ie.east) {...};
  \node[anchor=north, font=\fontsize{6.2}{7}\selectfont\sffamily\itshape, text=figgood] at ([yshift=-0.07in]iid.south) {item Semantic ID};
\end{scope}
\path[draw=figink, fill=figfaint, line width=0.7pt] (\PX, \BY) rectangle (\WB, \BY+\BH);
\draw[small flow] (\DXb, \BY+\BH/2) -- (\PX, \BY+\BH/2);
\node[inner sep=0, font=\fontsize{7}{8}\selectfont\sffamily\bfseries] at ({(\PX+\WB)/2}, \BY+\BH-0.13) {the page};
\foreach \r in {0,...,3} {
  \pgfmathsetmacro\ry{\BY+\BH-0.42-\r*0.28}
  \path[draw=figgpu, fill=white, line width=0.6pt] (\PX+0.05, \ry) rectangle ++(0.12, 0.13);
  \foreach \d in {0,...,4}
    {\path[draw=figgood, fill=white, line width=0.45pt] (\PX+0.205+\d*0.064, \ry+0.02) rectangle ++(0.048, 0.09);}
}
\pgfmathsetmacro\rx{\WA+3*(\BW+\G)+\BW/2}
\draw[figbad, line width=0.8pt, dash pattern=on 2.5pt off 1.6pt, -{Triangle[length=3pt, width=3pt]}]
  (\rx, \AY-0.02) -- (\rx, \BY+\BH-0.09);
\node[anchor=west, fill=white, inner sep=0.8pt, font=\fontsize{6.4}{7.5}\selectfont\sffamily\itshape, text=figbad]
  at (\rx+0.07, \AY-0.20) {rules move inside};
\end{tikzpicture}
\fi
\par\vspace{0.7ex}
{\footnotesize \hypertarget{fig:1}{}\textbf{Figure 1. From a ranking pipeline to one decode.} (a) The
pipeline ranks candidates, then applies the rules. (b) One decode writes
the page and checks the same rules at every token.\par}
\end{figure}

Generated Semantic IDs must name real candidates. Generative retrieval
\citep{tay2022dsi}, where a model emits one identifier token by token,
ensures this with a trie (prefix tree) over the real catalog that allows
only legal next tokens at each step \citep{decao2021genre, wang2022nci}.
A page, however, has rules that a fixed catalog trie cannot express, and
the same three recur across domains: an eligible set, the entries that
qualify for this request; a blocked set, the entries already output and
any they rule out; and nesting, one kind of entry inside another.

\begin{itemize}
\tightlist
\item
  Food delivery (this paper): stores, then menu items. A store is
  eligible if it delivers to the requester. A page holds at most one
  store per chain, and a store lists each item once.
\item
  E-commerce: sellers, then products. A product is eligible if it is in
  stock and ships to the address, and a page shows it at most once.
\item
  Video feeds: creators, then videos. A video is eligible if it is
  licensed in the viewer's region, and a feed shows it at most once.
\item
  Advertising: advertisers, then creatives. A creative is eligible while
  budget remains and targeting matches, and an advertiser stops
  appearing after it reaches its frequency cap.
\item
  Database queries: tables, then columns. A table is eligible if the
  user may read it, and a query selects each column at most once.
\end{itemize}

We call a constraint with these parts LEGIT, live eligibility gating
over identifier tries (\hyperlink{sec:2.4}{Section 2.4}). Prior work covers parts of this
problem. Methods that accelerate the trie walk
\citep{anandakumar2026flashtrie, su2026static} assume one catalog trie
shared across requests, GRACE \citep{fang2026grace} filters per request
but walks one kind of identifier, and ahead-of-time grammar engines
\citep{dong2025xgrammar, willard2023outlines, geng2025jsonschemabench}
assume a fixed grammar. None handles a constraint that changes with
every request and every token. We study one instance, the Uber Eats
search engine on its production catalog. Its decoder switches between a
store trie and an item trie, and three indexes (map cell to store, store
to item, store to chain) filter both tries. We built 36 implementations
of this constraint (\hyperlink{tab:2}{Table 2} names the main ones). The implementations
vary in where the trie walk executes, on the CPU or the GPU; how trie
nodes are laid out, padded to one width or ragged; and whether kernels
launch once per group of rows (sequences that need the same mask,
\hyperlink{sec:2.5}{Section 2.5}) or once per step. We measured them in two load regimes:
\emph{light}, where requests arrive slowly, so a decode step holds one
or two of them, and \emph{saturated}, where requests arrive faster than
pages finish, so many share each step.

\begin{figure*}[!tbp]
\centering
\iflatexml\includegraphics[width=448.9pt]{figures/decoder.svg}\else
\begin{tikzpicture}[fig, x=0.85in, y=0.82in]
\def\CXa{1.34}\def\CXb{6.10}\def\CYa{1.72}\def\CYb{4.70}
\def\SPLIT{3.06}
\def\BW{1.29}
\def\XA{1.495}\def\XB{3.075}\def\XC{4.655}   
\def\AY{3.28}\def\AH{0.62}   
\def\HY{2.02}\def\HH{0.80}   
\def\IY{0.10}\def\IH{1.36}   

\draw[figink, line width=0.9pt] (\CXa, \CYa) rectangle (\CXb, \CYb);
\draw[fignil, line width=0.6pt, dash pattern=on 3pt off 2.2pt] (\CXa, \SPLIT) -- (\CXb, \SPLIT);
\node[anchor=west, inner sep=0, font=\figlabel\itshape, text=figmuted] at (\CXa+0.09, \CYb-0.17) {page decoder, one step of the loop};
\node[anchor=east, inner sep=0, font=\figtitle, text=figgpu] at (\CXb-0.09, \CYb-0.17) {GPU};
\node[anchor=east, inner sep=0, font=\figtitle, text=figcpu] at (\CXb-0.09, \CYa+0.17) {CPU};

\FigBox[gpubox]{\XA}{\AY}{\BW}{\AH}{model forward\\and key value cache}
\FigBox[gpubox]{\XC}{\AY}{\BW}{\AH}{apply the mask,\\sample one token}
\draw[flow, figgpu] (\XA+\BW+0.02, \AY+\AH/2) -- (\XC-0.02, \AY+\AH/2)
  node[midway, above=1pt, inner sep=0, text=figgpu] {logits};
\draw[flow, figgpu] (\XC+\BW/2, \AY+\AH+0.03) to[bend right=17]
  node[midway, fill=white, inner sep=1.2pt, text=figgpu] {\ValMedianPageTokens\ tokens per page, median}
  (\XA+\BW/2, \AY+\AH+0.03);

\FigBox{\XA}{\HY}{\BW}{\HH}{group the rows\\by trie node, blocked\\set and grammar}
\FigBox[draw=figgood, line width=1.2pt]{\XB}{\HY}{\BW}{\HH}{walk the trie,\\build the legal\\child set}
\FigBox{\XC}{\HY}{\BW}{\HH}{scatter the legal\\set into a mask\\over the vocabulary}
\draw[flow] (\XB-0.27, \HY+\HH/2) -- (\XB-0.02, \HY+\HH/2);
\draw[flow] (\XC-0.27, \HY+\HH/2) -- (\XC-0.02, \HY+\HH/2);
\draw[thickflow, figcpu] (\XC+\BW/2, \HY+\HH+0.02) -- (\XC+\BW/2, \AY-0.02);
\node[anchor=west, fill=white, inner sep=1pt, text=figcpu] at (\XC+\BW/2+0.06, \HY+\HH+0.12) {mask};
\draw[figgood, line width=0.9pt, dash pattern=on 2.5pt off 1.6pt,
      {Triangle[length=3.6pt, width=3.6pt]}-{Triangle[length=3.6pt, width=3.6pt]}]
  (\XB+\BW/2, \HY+\HH+0.04) -- (\XB+\BW/2, \AY-0.04);
\node[anchor=east, align=right, fill=white, inner sep=1pt, font=\figtext\itshape, text=figgood]
  at (\XB+\BW/2-0.08, {(\HY+\HH+\AY)/2}) {this box can execute\\on either side};

\FigBox{0.02}{\AY}{1.10}{\AH}{\figsmall request\\\figsmall search query,\\\figsmall location as map cells}
\draw[flow] (1.14, \AY+\AH/2) -- (\CXa-0.02, \AY+\AH/2);
\FigBox{\CXb+0.24}{\AY}{0.94}{\AH}{\figsmall the page\\\figsmall 20 stores,\\\figsmall 5 items each}
\draw[flow] (\CXb+0.02, \AY+\AH/2) -- (\CXb+0.22, \AY+\AH/2);

\newcommand{\legalinput}[6]{%
  \FigBox[draw=#2, line width=0.9pt]{#1}{\IY}{\BW}{\IH}{}
  \node[anchor=north, inner sep=0, align=center, text width=\BW*0.85in-6pt, font=\figsmall]
    at ({#1+\BW/2}, \IY+\IH-0.10) {#3};
  \node[anchor=south, inner sep=0, align=center, text width=\BW*0.85in-6pt, font=\figsmall, text=figmuted]
    at ({#1+\BW/2}, \IY+0.32) {#4};
  \node[inner sep=0, font=\figtext\itshape, text=#2] at ({#1+\BW/2}, \IY+0.14) {#5};
  \draw[figink, line width=0.6pt, -{Triangle[length=3.2pt, width=3.2pt]}, draw=#2]
    ({#1+\BW/2}, \IY+\IH+0.02) -- ({\XB+\BW/2+#6}, \HY-0.02);}
\legalinput{\XA}{fignil}{store trie, item trie, store $\to$ item index, store $\to$ chain index}
  {$O(10^6)$ stores, $O(10^7)$ item identifiers, $O(10^9)$ store and item pairs}{fixed, built offline}{-0.34}
\legalinput{\XB}{figpurple}{map cell $\to$ store index: the request's map cells give eligible stores $E(h)$}
  {one sorted store list per map cell, at map resolutions 6 to 10}{changes per request}{0}
\legalinput{\XC}{figbad}{blocked set $S$: stores on the page and their chains, items in this store's row}
  {at most 20 stores, with their chains, and 5 items per row}{changes per token}{0.34}
\end{tikzpicture}
\fi
\par\vspace{0.7ex}
{\footnotesize \hypertarget{fig:2}{}\textbf{Figure 2. Where the constraint sits in the page decoder.} One
decode step. Everything that decides what goes into the mask is the
constraint. The trie walk can execute on either side of the boundary
between the CPU and the GPU. The three inputs at the bottom carry their
sizes in our benchmark and how often each changes. In the full CPU stack
(\hyperlink{sec:5.5}{Section 5.5}), the constraint box sends a short list of legal tokens,
and the GPU applies it to the logits.\par}
\end{figure*}

Our contributions are:

\begin{enumerate}
\def\labelenumi{\arabic{enumi}.}
\tightlist
\item
  A data design for mixed pages. One store trie and one global item trie
  hold the two kinds of identifier, and the page grammar, which fixes
  what kind of slot comes next, selects which one to walk. Every trie
  numbers its leaves depth first, so each rule becomes a sorted array
  and the legal-child test becomes a few binary searches. A
  store-to-item index supplies the eligible items of the store just
  emitted, so the design needs no trie per menu, and reversing the
  nesting needs no new offline artifact.
\item
  An ablation showing that the grammar alone keeps pages well formed,
  with every slot filled, but that without the trie and its indexes none
  of the generated stores exist.
\item
  Measurements of where the time goes. In every implementation the GPU's
  streaming multiprocessors (its compute units) are active at most 5.42
  percent of the time, while its utilization counter reads 12.3 to 27.8
  percent, because that counter treats the GPU as busy whenever any
  kernel runs. The GPU's compute units sit mostly idle, which points to
  launches and CPU synchronization as the cost.
\item
  A recipe that needs no custom kernel. Walking the trie on the CPU,
  with inner subtrie compaction (a small trie over each store's menu)
  and a cached blocked set per row, matches within noise our fastest GPU
  implementation, the full GPU stack, which launches once per step. We
  call this combination the CPU recipe. Three more changes give the full
  CPU stack, the fastest implementation on both page layouts we test,
  store first and item first: walking every group of rows while the
  model runs, answering trie queries with plain arrays, and copying all
  legal-token lists to the GPU in one transfer per step. The full CPU
  stack reduces a constrained token from about 5 times the cost of an
  unconstrained one to about 1.7 times.
\item
  A build order. We read the 36 implementations as a grid over
  placement, trie layout and launch structure. The same change can speed
  up one cell of the grid and slow down another. Ragged trie nodes, for
  example, speed up the CPU walk and slow down the GPU walk. From the
  grid we derive a benchmark method and a build order for teams
  implementing their own version.
\end{enumerate}

A separate appendix holds the full ranking, the catalog check of every
page and further checks.

\section[2. The constraint]{\hypertarget{sec:2}{}2. The constraint}\label{the-constraint}

\subsection[2.1 Pages of Semantic IDs]{\hypertarget{sec:2.1}{}2.1 Pages of Semantic IDs}\label{pages-of-semantic-ids}

We name each store and item by five discrete codes. Residual
quantization of its embedding gives the first three
\citep{rajput2023tiger, lee2022rq}, and optimized product quantization
of what remains gives the last two \citep{ge2014opq}. Level sizes are
{[}4096, 2048, 2048, 1024, 1024{]} for stores and {[}8192, 4096, 4096,
2048, 2048{]} for items. They describe about 2 × 10¹⁶ store tuples and 6
× 10¹⁷ item tuples, while the Uber Eats catalog uses only O(10⁶) store
and O(10⁷) item identifiers. A store's menu holds 64 items at the median
and 24,316 at the 99th percentile. Menu entries reduce to item
identifiers in two steps. Repeated text collapses O(10⁹) store and item
pairs to O(10⁸) distinct item texts, and similar texts then share an
identifier, about ten items per identifier. Only about one in 10¹⁰
well-formed identifiers names a real store or item, so output stays real
only through a lookup against the catalog, which we implement as a trie
whose edges are Semantic ID tokens. At each step the decoder reads the
children of the node matching the prefix and masks every other token
before sampling.

The model decides page length, and with twenty store slots the median
page is 843 tokens. \hyperlink{sec:5.6}{Sections 5.6} and \hyperlink{sec:5.7}{5.7} vary the layout and the slot
count. We initialize a 0.6 billion parameter decoder-only transformer
from Qwen3 \citep{yang2025qwen3} and serve it on vLLM
\citep{kwon2023vllm} with continuous batching on one NVIDIA B200 GPU per
benchmark run. The constraint is a custom logits processor (code that
edits the token scores before sampling). The forward pass is identical
across every implementation we compare.

\subsection[2.2 The system, end to end]{\hypertarget{sec:2.2}{}2.2 The system, end to end}\label{the-system-end-to-end}

As a running example, a person opens the app and searches for ``pizza''.
The request carries the query and the person's location, given as cells
of the H3 hexagonal map grid \citep{brodsky2018h3}. Production uses
resolutions 6 to 10. Our benchmark requests carry only the resolution-6
cell, so their eligible sets are smaller than in production. The model
writes one page of twenty stores, with five items under each store, and
we call a store's row of items a carousel. The page must follow four
rules. (1) Every store and every item exists in the catalog. (2) Every
store delivers to the person's location, and every item is on the menu
of the store above it. (3) No store appears twice on the page, and no
item appears twice in the same store's carousel. (4) At most one store
from each chain appears on the page. An independent store counts as a
chain of its own. Rule 1 holds for every request, rule 2 changes per
request, and rules 3 and 4 change with every identifier the page emits.
Each tag is one token and each identifier is five Semantic ID tokens, so
a page with one store and one item is

\begin{verbatim}
<page>
  <store> s1 s2 s3 s4 s5
    <item> i1 i2 i3 i4 i5 </item>
  </store>
</page>
\end{verbatim}

We build five artifacts offline, two tries and three indexes, and all
five share one numbering. We number the leaves of each trie in
depth-first order, so a store or an item is a single integer, its leaf
id.

\begin{itemize}
\tightlist
\item
  \textbf{Store trie and item trie.} Tries over the Semantic IDs of
  O(10⁶) stores and of the items of the stores that carry a menu. They
  enforce rule 1.
\item
  \textbf{Map-cell-to-store index.} For each map cell, the sorted leaf
  ids of the stores that deliver to it, built from each store's
  production delivery zone. It enforces rule 2 for stores.
\item
  \textbf{Store-to-item index.} For each store, the sorted leaf ids of
  its items, O(10⁹) store and item pairs in total, since menus share
  items. It enforces rule 2 for items.
\item
  \textbf{Store-to-chain index.} For each store, its chain, and for each
  chain, its stores. Of all stores, 37.5 percent belong to a parent
  chain. It enforces rule 4.
\end{itemize}

In a store slot, the legal tokens start from the union of the request's
cells in the map-cell-to-store index, then remove every store already on
the page and every store that shares a chain with one of them. In an
item slot, start from the menu of the store that owns the carousel, then
remove every item already in the carousel. In both, a child token of the
current trie node is legal if at least one remaining candidate lies
below it. Before the decoder opens a new slot, it checks at the trie
root that a candidate remains. If none does, it masks the opening tag,
so the decoder never starts an identifier it cannot finish. \hyperlink{sec:4}{Section 4}
shows how the shared numbering turns each test into a few binary
searches.

\subsection[2.3 The same rules
elsewhere]{\hypertarget{sec:2.3}{}2.3 The same rules
elsewhere}\label{the-same-rules-elsewhere}

A decoder may share our cost structure if an earlier identifier selects
the next trie, the request carries a filter, and an identifier may be
emitted only once. We did not study other applications. Table 15 in the
appendix maps the example onto six applications. If every filter is a
sorted array of leaf ids (\hyperlink{sec:4}{Section 4}), filters combine by intersection
and the range lookup applies unchanged.

\subsection[2.4 Why it is dynamic]{\hypertarget{sec:2.4}{}2.4 Why it is dynamic}\label{why-it-is-dynamic}

\textbf{Notation.} At decode step \texttt{t}, a sequence is in grammar
state \texttt{g} and sits at trie node \texttt{v}. The blocked set
\texttt{S} holds the stores already on the page with every store of
their chains, and the items already in the current carousel. The request
carries a set of map cells \texttt{h}, and \texttt{E(h)} is the set of
stores eligible in any of them. In an item slot, the menu of the store
that owns the carousel replaces \texttt{E(h)}. Outside an identifier,
the legal tokens are \texttt{A(g)}, the set the grammar alone allows.
Inside one, the grammar permits exactly the trie's tokens, so \texttt{g}
only selects which trie \texttt{v} belongs to.

Inside an identifier, the legal next tokens are

\[
\begin{aligned}
L(g, v, S, h) = \{\, c \in \mathrm{children}(v) :{}& \\
  \mathrm{subtree}(c) \cap (E(h) \setminus S) &\neq \emptyset \,\}.
\end{aligned}
\]

In words, a child token \texttt{c} is legal if at least one eligible
identifier that is not yet on the page lies below it in the trie. Three
properties follow, and together they rule out precompiling the
constraint.

\begin{enumerate}
\def\labelenumi{\arabic{enumi}.}
\tightlist
\item
  \texttt{L} depends on \texttt{h}, so it differs per request.
\item
  \texttt{L} depends on \texttt{S}, so it differs per token. Emitting an
  identifier shrinks the legal set for the rest of the page.
\item
  \texttt{L} is nested. The store the decoder chooses decides which
  items it may walk to next.
\end{enumerate}

Property 1 alone would permit a per-request compile. Property 2 rules
that out, and property 3 rules out listing the reachable states ahead of
time. We call a trie constraint with all three properties live
eligibility gating over identifier tries, or LEGIT. LEGIT names this
class of constraint, and the paper proposes no new decoding engine.

\subsection[2.5 Where the time goes]{\hypertarget{sec:2.5}{}2.5 Where the time goes}\label{where-the-time-goes}

The cost model below predicts that kernel launches and waits dominate
the cost of a step. A decode batch holds many rows, one per sequence
being written, at different positions. Rows that share a trie node and a
blocked set (compared by a hash of the set) need the same mask, so the
constraint sorts the batch into \texttt{G} groups. With \texttt{d} the
mean number of children at the nodes in play and \texttt{k} the cost of
one kernel launch (a call to the GPU) plus the CPU's wait for that
kernel to finish, a step costs

\[
C(\text{step}) = G \cdot k + \sum_{\text{groups}} O(d)
\]

The second term is the arithmetic, which stays small because a node has
tens of children on average and the vocabulary has tens of thousands of
tokens. The first term grows with \texttt{G} regardless of \texttt{d}.
\hyperlink{alg:1}{Algorithm 1} is the production loop that every implementation modifies.

\begin{algorithm}[!tbp]\hypertarget{alg:1}{}
\caption{The control.}
\small
\begin{algorithmic}[1]
\FOR{each decode step}
\STATE groups \(\gets\) partition rows by (node, blocked, state)
\item[] \hfill\(\triangleright\) G groups
\FOR[G iterations]{each group}
\STATE children \(\gets\) children of the node
\STATE legal \(\gets\) children minus blocked, in E(h)
\STATE mask \(\gets\) scatter legal into a fresh buffer \COMMENT{launch}
\IF[CPU read, waits]{mask is empty}
\STATE fall back to the grammar only set
\ENDIF
\STATE apply mask to the logits \COMMENT{launch}
\ENDFOR
\STATE sample
\ENDFOR
\end{algorithmic}
\end{algorithm}

\section[3. The cost of valid pages]{\hypertarget{sec:3}{}3. The cost of valid pages}\label{the-cost-of-valid-pages}

The constraint is expensive, and valid pages depend on it. The control
in our tables, a direct GPU implementation that walks a padded trie on
the GPU, makes each token about 5 times as slow as unconstrained
decoding at light load. The full CPU stack brings that to about 1.7
times (\hyperlink{sec:5.5}{Section 5.5}). \hyperlink{tab:1}{Table 1} removes one restriction at a time and
checks the pages produced. Dropping the trie, the request indexes and
the blocked set together saves 63.4 percent, but none of the resulting
pages' stores exist.

\begin{table*}[!tbp]
\centering
{\footnotesize \hypertarget{tab:1}{}\textbf{Table 1.} What each restriction costs and what removing it
breaks. Each row removes one restriction and keeps the rest, and the
last two rows also remove every restriction above them. The saved
columns give the percent of the control's time per token that the
removal saves, over six rounds per regime. The request indexes are the
map-cell-to-store and store-to-item indexes. Pages stay well formed on
every row but the last.\par}
\vspace{0.7ex}
\footnotesize
\begin{adjustbox}{max width=\linewidth}
\begin{tabular}{@{}l l l r r l l@{}}
\toprule
\thead{restriction removed} & \thead{saved, light} & \thead{saved, saturated} & \thead{ms/token, light} & \thead{tokens/s, light} & \thead{invalid entries emitted} & \thead{what breaks} \\
\midrule
none, the control & 0 (reference) & 0 (reference) & 10.22 & 98 & none &
nothing \\
request indexes & 12.4\% & 29.7\% & 8.94 & 112 & 99.7\% of stores,
100.0\% of items & out of reach, off the menu \\
blocked set & 25.6\% & 55.1\% & 7.61 & 131 & 82.5\% of stores are
repeats & 97\% of pages repeat a store \\
trie, indexes and blocked set & 63.4\% & 97.7\% & 3.74 & 267 & 100\% of
stores are nonexistent & pages still parse \\
everything, grammar too & 79.5\% & 98.9\% & 2.09 & 479 & not applicable
& no page parses \\
\bottomrule
\end{tabular}
\end{adjustbox}
\end{table*}

The trie row of \hyperlink{tab:1}{Table 1} shows why the constraint is needed. Pages
decoded without the trie and its indexes still parse and fill all twenty
slots, and only 0.3 percent repeat a store, because nonexistent
identifiers rarely collide. Yet none of their stores exist, and nothing
in their shape shows it. Grammar engines guarantee well-formed output,
which our grammar alone already provides. Guaranteeing that every
identifier names a real entry the request can reach accounts for most of
the cost.

\section[4. The trie and index design]{\hypertarget{sec:4}{}4. The trie and index design}\label{the-trie-and-index-design}

Every rule reduces to one range lookup. The cost model leaves out one
step: for each candidate child \texttt{c}, a test of whether
\texttt{subtree(c)\ intersects\ E(h)\ \textbackslash{}\ S}, that is,
whether the child's subtree holds an eligible leaf not yet blocked. As a
set intersection per child, this test would dominate the loop. With the
design below the test costs a few binary searches, so its arithmetic
stays small and launches and waits make up the rest of the cost. \hyperlink{fig:3}{Figure
3} shows the design on a toy trie.

\begin{figure*}[!tbp]
\centering
\iflatexml\includegraphics[width=400.6pt]{figures/structures.svg}\else
\begin{tikzpicture}[fig, x=0.74in, y=0.74in]
\def\H{4.02}
\newcommand{\mono}{\fontsize{7}{8}\selectfont\ttfamily}
\newcommand{\tnode}[5][]{%
  \path[draw=figink, fill=white, line width=0.8pt, #1] ({#2}, {#3}) circle ({#4});
  \node[inner sep=0, font=\figsmall] at ({#2}, {#3}) {#5};}
\newcommand{\tedge}[6]{%
  \draw[figink, line width=0.6pt] ({#1}, {#2-0.19}) -- ({#3}, {#4+#5})
    node[pos=0.58, fill=white, inner sep=0.8pt, font=\figlabel\itshape, text=figgpu] {#6};}

\node[font=\figtitle, anchor=west, inner sep=0] at (0.05, \H-0.14) {A\enspace A trie, leaves numbered depth first};
\pgfmathsetmacro\RY{\H-0.58}
\pgfmathsetmacro\MY{\H-1.28}
\pgfmathsetmacro\LY{\H-1.98}
\tedge{1.55}{\RY}{0.85}{\MY}{0.19}{a}
\tedge{1.55}{\RY}{2.35}{\MY}{0.19}{b}
\tnode{1.55}{\RY}{0.19}{[0,5)}
\tnode{0.85}{\MY}{0.19}{[0,3)}
\tnode{2.35}{\MY}{0.19}{[3,5)}
\foreach \i/\x/\px/\tok/\e/\s in {0/0.30/0.85/x/0/0, 1/0.85/0.85/y/1/1, 2/1.40/0.85/z/1/1,
                                  3/2.05/2.35/x/0/0, 4/2.65/2.35/y/1/0} {
  \tedge{\px}{\MY}{\x}{\LY}{0.17}{\tok}
  \ifnum\e=1 \tnode[draw=figgpu, line width=1.3pt]{\x}{\LY}{0.17}{}%
  \else \tnode[draw=fignil, line width=0.7pt]{\x}{\LY}{0.17}{}\fi
  \ifnum\s=1
    \draw[figbad, line width=1.3pt] (\x-0.12, \LY-0.12) -- (\x+0.12, \LY+0.12) (\x-0.12, \LY+0.12) -- (\x+0.12, \LY-0.12);
  \fi
  \node[inner sep=0, font=\figsmall] at (\x, \LY-0.29) {leaf \i};
}
\FigNote[text=figgpu]{0.05}{\LY-0.56}{blue ring: eligible for this request, E = \{1, 2, 4\}}
\FigNote[text=figbad]{0.05}{\LY-0.76}{red cross: already on this page, S = \{1, 2\}}
\FigNote{0.05}{\LY-0.96}{in each node: its leaf range [node\_lo, node\_hi)}

\def\XA{3.35}
\def\XO{4.30}
\def\CW{0.33}
\def\CH{0.24}
\node[font=\figtitle, anchor=west, inner sep=0] at (\XA, \H-0.14) {B\enspace Stored as compressed sparse row arrays};
\newcommand{\arr}[5]{%
  \node[anchor=east, inner sep=0, font=\mono] at (\XO-0.08, {#1}) {#2};
  \foreach \v [count=\c from 0] in {#3} {%
    \pgfmathtruncatemacro\lit{\c>=#4 && \c<=#5}%
    \ifnum\lit=1 \FigCell[draw=figgpu, fill=figtint]{\XO+\c*\CW}{#1}{\CW}{\CH}{\v}%
    \else \FigCell{\XO+\c*\CW}{#1}{\CW}{\CH}{\v}\fi}}
\FigNote[font=\figtext\itshape]{\XA}{\H-0.42}{per node (nodes numbered breadth first)}
\arr{\H-0.69}{node}{0,1,2,3,4,5,6,7}{9}{-1}
\arr{\H-0.98}{node\_lo}{0,0,3,0,1,2,3,4}{9}{-1}
\arr{\H-1.27}{node\_hi}{5,3,5,1,2,3,4,5}{9}{-1}
\arr{\H-1.56}{offsets}{0,2,5,7,7,7,7,7,7}{0}{1}
\FigNote[font=\figtext\itshape]{\XA}{\H-1.84}{per edge}
\arr{\H-2.11}{edge}{0,1,2,3,4,5,6}{9}{-1}
\arr{\H-2.40}{tokens}{a,b,x,y,z,x,y}{0}{1}
\arr{\H-2.69}{child}{1,2,3,4,5,6,7}{0}{1}
\FigNote[text=figgpu]{\XA}{\H-3.00}{The root's children are edges offsets[0] to offsets[1] $-$ 1\\(edges 0 and 1), one slice.}

\pgfmathsetmacro\TY{\H-3.36}
\node[font=\figtitle, anchor=west, inner sep=0] at (0.05, \TY) {C\enspace Legal children of the root: a child is legal when its eligible leaves in range outnumber its blocked ones};
\foreach \x/\head in {0.10/child, 0.85/{[node\_lo, node\_hi)}, 2.30/eligible in range, 3.70/blocked in range, 5.00/legal}
  {\FigNote{\x}{\TY-0.24}{\head}}
\foreach \k/\tok/\rng/\e/\s/\ok/\col in {0/a/{[0,3)}/2/2/no/figbad, 1/b/{[3,5)}/1/0/yes/figgood} {
  \pgfmathsetmacro\y{\TY-0.46-0.20*\k}
  \foreach \x/\v in {0.10/\tok, 0.85/\rng, 2.30/\e, 3.70/\s}
    {\FigNote[text=figink, font=\mono]{\x}{\y}{\v}}
  \FigNote[text=\col, font=\mono]{5.00}{\y}{\ok}
}
\FigNote{5.55}{\TY-0.56}{a is refused although\\the request reaches it:\\its eligible leaves are all\\blocked, so no dead end.}
\end{tikzpicture}
\fi
\par\vspace{0.7ex}
{\footnotesize \hypertarget{fig:3}{}\textbf{Figure 3. The structures, on a five-leaf trie.} Panel A numbers
the leaves depth first, so the leaves under any node form one range,
which the node stores as \texttt{node\_lo} and \texttt{node\_hi}. Panel
B shows how we store the trie: per-node arrays hold each node's leaf
range and \texttt{offsets}, and per-edge arrays hold each edge's token
and \texttt{child} node. Node \texttt{v}'s children are edges
\texttt{offsets{[}v{]}} to \texttt{offsets{[}v\ +\ 1{]}\ -\ 1}, one
slice. Panel C runs the legal-child test at the root for a request whose
eligible leaves are 1, 2 and 4, on a page that has already used 1 and 2.
The test refuses prefix a: the request can reach it, but every eligible
leaf under it is already blocked.\par}
\end{figure*}

\begin{itemize}
\tightlist
\item
  We store each trie in compressed sparse row form. A node's children
  are one contiguous slice, so reading a child list takes one slice and
  masking takes that slice plus a scatter. Each trie has one leaf per
  distinct identifier, O(10⁶) leaves in 0.16 gigabytes for stores and
  O(10⁷) in 4.0 gigabytes for items. The item trie is global, since
  building a trie for each of O(10⁶) menus is not feasible.
\item
  We number leaves depth first, so the leaves under any node form one
  range, \texttt{{[}node\_lo,\ node\_hi)}. The eligible set arrives as a
  sorted array of leaf ids, so testing a subtree against it is two
  binary searches.
\item
  A child is legal when its range holds more eligible leaves than
  blocked ones, so excluding blocked entries is a subtraction. With no
  filter the eligible count is \texttt{node\_hi\ -\ node\_lo}. Counting
  before descending keeps the decoder out of any prefix whose
  completions are all blocked, which would leave an empty legal set
  mid-identifier.
\item
  A leaf id is the rank of an identifier in sorted order, and depends on
  nothing else. We can build the trie, the map-cell-to-store index and
  the store-to-item index independently, and they line up when built
  from the same catalog snapshot. Every artifact records the leaf count
  we built it against, and the runtime refuses a mismatch. Equal counts
  alone cannot prove the same snapshot. A fingerprint of the sorted
  identifiers would be a safer check, which we did not implement.
\item
  The store-to-item index is a second compressed sparse row structure,
  about 11 gigabytes of 32-bit integers in CPU memory, addressed by
  store leaf id.
\end{itemize}

\hyperlink{alg:2}{Algorithm 2} is the whole of the mask arithmetic. Every implementation in
\hyperlink{sec:5}{Section 5} computes exactly these masks and differs only in where
\hyperlink{alg:2}{Algorithm 2} executes and how many launches that takes.

\begin{algorithm}[!tbp]\hypertarget{alg:2}{}
\caption{The legal-child test. \texttt{E} is the sorted
eligible leaf ids and \texttt{S} the sorted blocked ones inside
\texttt{E}.}
\small
\begin{algorithmic}[1]
\FUNCTION{\textsc{legal\_children}(v, E, S)}
\STATE legal \(\gets\) empty
\FOR[one slice]{edge in offsets[v] .. offsets[v + 1] - 1}
\STATE c \(\gets\) child[edge]
\STATE lo, hi \(\gets\) node\_lo[c], node\_hi[c] \COMMENT{leaves of c}
\STATE avail \(\gets\) count of E in [lo, hi) \COMMENT{two searches}
\STATE blocked \(\gets\) count of S in [lo, hi) \COMMENT{two searches}
\IF[an unblocked leaf left]{avail > blocked}
\STATE add tokens[edge] to legal
\ENDIF
\ENDFOR
\STATE \textbf{return} legal
\ENDFUNCTION
\end{algorithmic}
\end{algorithm}

Compressed sparse row tries and depth-first interval labels are both
old, and STATIC uses the same trie layout. Our contribution is their
combination under this constraint. One leaf numbering, shared by
artifacts that never communicate, makes the per-request test a range
lookup. A blocked count then reduces the rule that no entry repeats to
one comparison per candidate and removes dead ends.

\section[5. Results]{\hypertarget{sec:5}{}5. Results}\label{results}

\subsection[5.1 Setup]{\hypertarget{sec:5.1}{}5.1 Setup}\label{setup}

\begin{itemize}
\tightlist
\item
  Requests are sampled from one week of logged Uber Eats searches, and
  the logs fall under Uber's existing production data retention and
  anonymization policies. The prompt holds no user or device identifier.
\item
  Our metric is time per token: end-to-end request time, queueing
  included, divided by the output tokens of the page returned. We sample
  three candidates per page, so we cannot record time to first token or
  separate queue wait from decode time.
\item
  We treat the run as the unit of analysis. We collapse each run to its
  median and take intervals from variation across rounds. Requests in
  one run share a process and a cache, and pooling them would shrink
  intervals by about a factor of ten without adding information.
\item
  We run one round at a time. In each round every implementation runs
  once, all in flight together, with the order reshuffled so queue
  position is not confounded with implementation. Every comparison is
  against the control of the same round, so drift between rounds
  cancels.
\item
  We measure two load regimes, because they disagree often enough that
  quoting one alone would mislead. Light: requests arrive every 6.7
  seconds (0.15 queries per second) and a control page takes about 9.1
  seconds, so about 1.4 requests are in flight, fewer for faster
  implementations. The light-load figure measures the decoder.
  Saturated: requests arrive at 0.7 per second, faster than pages
  finish, so each request spends most of its time in the queue. A change
  here measures how fast an implementation drains a full batch, which is
  closer to throughput than to the cost of one decode step, and depends
  on run length. Appendix D gives pages per second.
\item
  A percentage is the change in time per token against the control, so a
  negative value is faster. Throughput and \hyperlink{tab:1}{Table 1}'s savings use the
  opposite sign. Percentage changes do not add, so to split a result we
  report speedups (control time divided by implementation time), which
  multiply.
\item
  We ran 36 implementations in nine sweeps: 1,258 runs and 125,800
  pages. A run is one implementation in one round and one regime. Every
  implementation carries 12 rounds per regime, and the control 60.
  Appendix A adds up these counts, Appendix B finds that 61 of the 70
  ranking tests survive a false discovery rate correction, and Appendix
  C checks that no implementation changes its pages.
\end{itemize}

\hyperlink{tab:2}{Table 2} names the implementations the text discusses, and Table 9 ranks
every option.

\begin{table}[!tbp]
\centering
{\footnotesize \hypertarget{tab:2}{}\textbf{Table 2.} The named implementations. Each row adds its options
to the row above. Launch is how often the trie walk launches kernels.
Overlap walks every group during the forward pass, and the numpy trie
answers trie queries with numpy on plain arrays. All compute the same
legal set and pages (Appendix C).\par}
\vspace{0.7ex}
\begin{adjustbox}{max width=\linewidth}
\begin{tabular}{@{}lll@{}}
\toprule
name & adds to the row above & launch \\
\midrule
\emph{GPU, padded} & & \\
control & \hyperlink{alg:1}{Algorithm 1} & group \\
fused kernel & one kernel per group & group \\
fused stack & batched launch, deferred check & step \\
full GPU stack & blocked set memo & step \\
\emph{CPU, ragged} & & \\
CPU trie & the walk, on the CPU & none \\
CPU recipe & inner subtrie compaction, memo & none \\
full CPU stack & overlap, numpy trie, batched write & none \\
\bottomrule
\end{tabular}
\end{adjustbox}
\end{table}

\subsection[5.2 Trie layout and walk
placement]{\hypertarget{sec:5.2}{}5.2 Trie layout and walk
placement}\label{trie-layout-and-walk-placement}

The 36 implementations are points in a grid with three axes: where the
trie walk executes, whether the node structure is padded or ragged, and
whether the walk pays for its launches per group or per step. The third
axis exists only on the GPU, since the CPU walk launches no kernels.
Each placement has a natural layout. The GPU runs threads in lockstep,
so padding every node to one width suits it. The CPU walks one node at a
time, so it gains from ragged children with no padding to skip. \hyperlink{tab:3}{Table 3}
crosses layout and placement. It holds the padded GPU control, the
ragged CPU walk, and two mismatched ablation cells.

\begin{table*}[!tbp]
\centering
{\footnotesize \hypertarget{tab:3}{}\textbf{Table 3.} Time per token at light load for each trie layout and
walk placement, as percent change against the padded GPU control, with
95 percent intervals over twelve rounds. Bold marks the fastest cell.
Saturated, the ragged GPU, padded CPU and ragged CPU cells are +108.9,
−10.3 and −40.7 percent. Pooled values in \hyperlink{tab:4}{Tables 4} and 9 differ by up to
2.5 points.\par}
\vspace{0.7ex}
\footnotesize
\begin{adjustbox}{max width=\linewidth}
\begin{tabular}{@{}l l l@{}}
\toprule
\thead{} & \thead{padded layout} & \thead{ragged layout} \\
\midrule
\textbf{on the GPU} & 0\% (control, the reference) & +113.4\% {[}+98.0,
+128.9{]}, 12 rounds \\
\textbf{on the CPU} & +4.2\% {[}+0.1, +8.3{]}, 12 rounds &
\textbf{−9.7\%} {[}−13.1, −6.2{]}, 12 rounds \\
\bottomrule
\end{tabular}
\end{adjustbox}
\end{table*}

When saturated the three non-control cells keep the same order, but the
padded CPU cell turns from a 4.2 percent loss into a 10.3 percent gain.
Padding fails on the CPU because it gives every node a fixed 8,192
slots, so a CPU walk reads mostly padding (Figure 8, Appendix J). On the
GPU, ragged children more than double the cost. The two axes interact
strongly. The interaction (the effect of ragged against padded on the
CPU, minus the same effect on the GPU) has magnitude 127.3 percentage
points at light load, against main effects of 59.4 for placement and
49.8 for layout.

\subsection[5.3 Effects that depend on
placement]{\hypertarget{sec:5.3}{}5.3 Effects that depend on
placement}\label{effects-that-depend-on-placement}

\begin{itemize}
\tightlist
\item
  Inner subtrie compaction builds a small trie over one store's eligible
  items. When saturated it makes the CPU walk 22.3 percent faster, in
  all twelve rounds, and the GPU walk 9.1 percent slower, a cost that is
  not significant after correction.
\item
  Batching the group launches improves the fused kernel by 10.5 points
  at light load and by 31.2 when saturated (Table 9, Appendix B).
\end{itemize}

\subsection[5.4 The ranking]{\hypertarget{sec:5.4}{}5.4 The ranking}\label{the-ranking}

\hyperlink{tab:4}{Table 4} shows the leaders and the rows the text discusses, and Table 9
in the appendix ranks the 35 implementations other than the control.

\begin{table*}[!tbp]
\centering
{\footnotesize \hypertarget{tab:4}{}\textbf{Table 4.} Selected rows of the ranking, as percent change in
time per token against the control, pooled across sweeps by
inverse-variance weight, with 95 percent intervals (n.s. marks one
spanning zero). The full CPU stack is composed through the overlap row
(Appendix K). Tokens per second is for one request at light load. Bold
marks the best point estimate in each group, which need not be a
significant lead. The bold saturated CPU cell mostly reflects the
control's queue (\hyperlink{sec:5.1}{Section 5.1}).\par}
\vspace{0.7ex}
\footnotesize
\begin{adjustbox}{max width=\linewidth}
\begin{tabular}{@{}l l l r@{}}
\toprule
\thead{implementation} & \thead{light} & \thead{saturated} & \thead{tokens/s, light} \\
\midrule
control: padded trie on the GPU & 10.56 ms, the reference & 194.43 ms,
the reference & 95 \\
\emph{walk on the GPU} & & & \\
inner subtrie compaction & +2.2\% {[}−1.1, +5.6{]}, n.s. & +9.1\%
{[}+3.0, +15.3{]} & 93 \\
fused kernel & −8.9\% {[}−11.9, −6.0{]} & −11.0\% {[}−17.1, −4.9{]} &
104 \\
fused stack & −16.8\% {[}−20.1, −13.6{]} & −49.8\% {[}−53.9, −45.6{]} &
114 \\
full GPU stack & \textbf{−21.6\% {[}−24.8, −18.3{]}} & \textbf{−56.8\%
{[}−61.1, −52.4{]}} & \textbf{121} \\
\emph{walk on the CPU} & & & \\
CPU trie & −10.0\% {[}−13.3, −6.6{]} & −41.2\% {[}−45.2, −37.2{]} &
105 \\
CPU trie + inner subtrie compaction & −16.7\% {[}−20.3, −13.1{]} &
−55.1\% {[}−60.3, −49.9{]} & 114 \\
CPU recipe & −20.2\% {[}−22.5, −17.8{]} & −63.5\% {[}−69.0, −58.0{]} &
119 \\
CPU recipe + overlap & −34.7\% {[}−37.0, −32.3{]} & −77.6\% {[}−82.9,
−72.3{]} & 145 \\
full CPU stack & \textbf{−65.9\% {[}−67.5, −64.3{]}} & \textbf{−97.9\%
{[}−98.2, −97.5{]}} & \textbf{278} \\
\emph{ablation} & & & \\
replay the prefix each step & +53.9\% {[}+49.7, +58.0{]} & +91.1\%
{[}+83.8, +98.4{]} & 62 \\
\bottomrule
\end{tabular}
\end{adjustbox}
\end{table*}

The full GPU stack and the CPU recipe differ by 0.7 standard errors at
light load and 1.9 when saturated, so twelve rounds cannot tell them
apart. We do not test equivalence. The full CPU stack beats both. Its
estimate is composed through the overlap row (Appendix K), since it
never ran paired with the full GPU stack.

\subsection[5.5 Explaining the fastest
implementations]{\hypertarget{sec:5.5}{}5.5 Explaining the fastest
implementations}\label{explaining-the-fastest-implementations}

Every batched GPU implementation, from the fused kernel with batched
launches up to the full GPU stack, pays the launch cost once per step.
\hyperlink{alg:3}{Algorithm 3} computes the same masks as \hyperlink{alg:1}{Algorithm 1} and changes only how
many times the step pays the \(G \cdot k\) term.

\begin{table*}[!tbp]
\centering
{\footnotesize \hypertarget{tab:5}{}\textbf{Table 5.} Hardware counters by walk placement, averaged over 24
GPU (control included) and 12 CPU implementations at light load.
Utilization is the share of time any kernel runs. Counters include idle
gaps, so they bound compute activity, and differences under a point are
noise.\par}
\vspace{0.7ex}
\begin{adjustbox}{max width=\linewidth}
\begin{tabular}{@{}llll@{}}
\toprule
counter & GPU walk & CPU walk & families overlap \\
\midrule
utilization, percent & 24.10 & 13.77 & no \\
multiprocessor active, percent & 4.67 & 4.76 & yes \\
multiprocessor occupancy, percent & 0.76 & 0.77 & yes \\
CPU cores used, of 4 & 1.58 & 2.51 & yes \\
\bottomrule
\end{tabular}
\end{adjustbox}
\end{table*}

\begin{algorithm}[!tbp]\hypertarget{alg:3}{}
\caption{The full GPU stack, the fastest GPU
implementation.}
\small
\begin{algorithmic}[1]
\STATE memo \(\gets\) each row's blocked set, reset per page
\item[]
\FOR{each decode step}
\STATE refresh memo[row] if its slot just closed \COMMENT{memo}
\STATE groups \(\gets\) partition rows by (node, blocked, state)
\STATE masks \(\gets\) one fused launch for ALL groups \COMMENT{batched}
\STATE apply masks to the batch \COMMENT{one launch}
\STATE record emptiness in the buffer, read it next step \COMMENT{defer}
\STATE sample
\ENDFOR
\end{algorithmic}
\end{algorithm}

The batched launch removes the per-group launch, the deferred check
removes the per-group CPU synchronization, and the memo removes a
per-row sort of the blocked set from every step that does not close a
slot. A CUDA graph over the mask step did not improve on the full GPU
stack (Table 9).

\begin{algorithm*}[!tbp]\hypertarget{alg:4}{}
\caption{The full CPU stack: the CPU recipe plus overlap,
numpy trie and batched write.}
\small
\begin{algorithmic}[1]
\STATE memo \(\gets\) each row's blocked set; cache \(\gets\) trie state per prefix \COMMENT{memo, numpy trie}
\FOR[the forward is still running]{each decode step}
\STATE refresh memo[row] if its slot just closed
\FOR[CPU only, no GPU op (overlap)]{each group of rows}
\STATE state \(\gets\) cache[prefix], else walk and store it
\STATE legal \(\gets\) \hyperlink{alg:2}{Algorithm 2} over the child slice \COMMENT{a token list}
\ENDFOR
\STATE one pinned copy of every group's list, not waited for \COMMENT{batched write}
\STATE apply the lists to the logits, then sample \COMMENT{about five launches}
\ENDFOR
\end{algorithmic}
\end{algorithm*}

\begin{figure*}[!tbp]
\centering
\iflatexml\includegraphics[width=393.2pt]{figures/maskways.svg}\else
%
\begin{tikzpicture}[fig, x=0.74in, y=0.62in]
\def\G{3}         
\def\LH{0.27}     
\def\XO{0.62}     
\def\S{2.35}      
\pgfmathsetmacro\FWD{0.55*\S}   
\def\W{7.3}

\newcommand{\figseg}[5]{%
  \path[#4] ({#1}, {#2+0.015}) rectangle ++({#3}, {\LH-0.03});
  \node[font=\figsmall, inner sep=0] at ({#1+(#3)/2}, {#2+\LH/2}) {#5};}
\newcommand{\lanes}[3]{%
  \pgfmathsetmacro\yd{#1}\pgfmathsetmacro\yh{#1+\LH+0.06}%
  \node[font=\figtitle, anchor=west, inner sep=0] at (0.05, #1+2*\LH+#3) {#2};
  \node[font=\figlabel, text=figcpu, anchor=east, inner sep=0] at (\XO-0.08, \yh+\LH/2) {CPU};
  \node[font=\figlabel, text=figgpu, anchor=east, inner sep=0] at (\XO-0.08, \yd+\LH/2) {GPU};
  \draw[lane] (\XO, \yh) -- (\W-0.05, \yh);
  \draw[lane] (\XO, \yd) -- (\W-0.05, \yd);}
\newcommand{\readback}[1]{\draw[readback] ({#1}, {\yd+\LH-0.01}) -- ({#1}, {\yh+0.02});}
\newcommand{\stepend}[1]{%
  \draw[stepend] ({#1}, {\yd-0.02}) -- ({#1}, {\yh+\LH});
  \node[font=\figtext, text=figmuted, anchor=west, inner sep=0] at ({#1+0.05}, {\yh+\LH/2}) {step ends};}

\lanes{4.05}{(a) Control: walk on the GPU, a launch per group}{0.44}
\figseg{\XO}{\yd}{\FWD}{seg=fignil}{\color{white}model forward}
\foreach \g in {0,...,2} {
  \foreach \j/\kw in {0/0.07, 1/0.07, 2/0.13} {
    \pgfmathsetmacro\t{\XO+\FWD+\g*0.63*\S+\j*0.18*\S}
    \pgfmathsetmacro\k{\kw*\S}
    \figseg{\t}{\yh}{0.05*\S}{seg=figlaunch}{}
    \figseg{\t+0.05*\S}{\yd}{\k}{seg=figgpu}{}
    \figseg{\t+0.05*\S}{\yh}{\k+0.04*\S}{hatched}{}
    \readback{\t+0.07*\S+\k}
  }
}
\pgfmathsetmacro\gA{\XO+\FWD}\pgfmathsetmacro\gB{\XO+\FWD+0.60*\S}
\draw[figink, line width=0.5pt, {Bar[width=4pt]}-{Bar[width=4pt]}] (\gA, \yh+\LH+0.08) -- (\gB, \yh+\LH+0.08);
\node[anchor=south, inner sep=0] at ($({(\gA+\gB)/2}, \yh+\LH+0.12)$) {one group};
\node[anchor=south west, inner sep=0, text=figmuted] at (\gB+0.08, \yh+\LH+0.12) {repeated for every group, \G\ here};
\pgfmathsetmacro\t{\XO+\FWD+3*0.63*\S}
\figseg{\t}{\yd}{0.16*\S}{seg=figgpu}{\color{white}apply}
\stepend{\t+0.16*\S}

\lanes{2.80}{(b) Full GPU stack: a fixed number of launches per step, no waits}{0.28}
\figseg{\XO}{\yd}{\FWD}{seg=fignil}{\color{white}model forward}
\pgfmathsetmacro\t{\XO+\FWD}
\figseg{\t}{\yh}{0.34*\S}{waitfill}{group rows}
\pgfmathsetmacro\t{\t+0.34*\S}
\figseg{\t}{\yh}{0.05*\S}{seg=figlaunch}{}
\figseg{\t+0.05*\S}{\yd}{0.40*\S}{seg=figgpu}{\color{white}all \G\ groups}
\pgfmathsetmacro\t{\t+0.45*\S+0.03}
\figseg{\t}{\yd}{0.16*\S}{seg=figgpu}{\color{white}apply}
\pgfmathsetmacro\t{\t+0.16*\S}
\stepend{\t}
\draw[figgood, line width=0.7pt, dash pattern=on 2.5pt off 1.5pt, -{Triangle[length=3.2pt, width=3.2pt]}]
  (\t-0.1, \yd+\LH) to[bend right=25] (\t+0.55, \yd+\LH/2);
\node[anchor=west, inner sep=0, text=figgood] at (\t+0.60, \yd+\LH/2) {empty check read next step};

\lanes{1.66}{(c) CPU recipe: walk on the CPU, a blocking copy per group}{0.28}
\figseg{\XO}{\yd}{\FWD}{seg=fignil}{\color{white}model forward}
\figseg{\XO+0.04*\S}{\yh}{\FWD-0.04*\S}{waitfill}{waits for the forward}
\foreach \g in {0,...,2} {
  \pgfmathsetmacro\t{\XO+\FWD+\g*(0.06+0.24*\S)}
  \readback{\t}
  \figseg{\t+0.01}{\yh}{0.19*\S}{seg=figcpu}{\color{white}walk \the\numexpr\g+1\relax}
  \pgfmathsetmacro\u{\t+0.01+0.19*\S}
  \draw[copyto] (\u+0.03, \yh+0.01) -- (\u+0.03, \yd+\LH-0.01);
  \figseg{\u+0.03}{\yd}{0.05*\S}{seg=figgpu}{}
  \ifnum\g<2 \figseg{\u}{\yh}{0.05+0.05*\S}{hatched}{}\fi
}
\stepend{\XO+\FWD+3*(0.06+0.24*\S)}

\lanes{0.58}{(d) Full CPU stack: walk during the forward, numpy trie, one batched write}{0.28}
\figseg{\XO}{\yd}{\FWD}{seg=fignil}{\color{white}model forward}
\foreach \g in {0,...,2} {
  \figseg{\XO+0.04*\S+\g*0.08*\S}{\yh}{0.07*\S}{seg=figcpu}{\color{white}\the\numexpr\g+1\relax}
}
\draw[flow] (\XO+0.28*\S, \yh+0.01) -- (\XO+\FWD+0.02, \yd+\LH-0.01);
\pgfmathsetmacro\t{\XO+\FWD+0.02}
\figseg{\t}{\yd}{0.15*\S}{seg=figgpu}{\color{white}write}
\stepend{\t+0.15*\S}
\node[anchor=west, inner sep=0, text=figgood] at (\t+0.15*\S+1.0, \yh+\LH/2) {walks hidden under the forward};

\begin{scope}[shift={(0.05, 0.25)}]
  \newcommand{\key}[3]{\path[#2] (#1, 0.01) rectangle ++(0.17, 0.12);
    \node[anchor=west, inner sep=0] at (#1+0.22, 0.07) {#3};}
  \key{0}{seg=figlaunch}{CPU queues a kernel}
  \key{1.85}{seg=figgpu}{GPU kernel}
  \key{3.25}{hatched}{CPU waits}
  \key{4.6}{seg=figcpu}{CPU walk}
\end{scope}
\draw[readback] (0.13, -0.01) -- (0.13, 0.15);
\node[anchor=west, inner sep=0] at (0.27, 0.07) {read back: the CPU waits for the GPU};
\draw[copyto] (3.13, 0.15) -- (3.13, -0.01);
\node[anchor=west, inner sep=0] at (3.27, 0.07) {copy from the CPU to the GPU};
\end{tikzpicture}
\fi
\par\vspace{0.7ex}
{\footnotesize \hypertarget{fig:4}{}\textbf{Figure 4. One decode step, four ways to build the mask.}
Schematic, with three groups of two prefix tokens each. Each panel title
names the implementation and how it launches (Appendix K).\par}
\end{figure*}

\hyperlink{fig:4}{Figure 4} draws one step four ways. On the GPU, one group's mask is a
chain of small kernels. Each read-back makes the CPU wait for the GPU's
queue to drain, and the GPU then waits for the next launch. On the CPU,
the walk launches no kernel chain and copies one mask per group. The GPU
path pays a cost once per group, and the CPU walk removes it. If this
per-group cost is the cause, the saving should grow with the number of
groups per step, and the CPU walk is indeed 9.7 percent faster at light
load and 40.7 percent faster when saturated, where steps hold more
groups (the saturated figure also includes queueing, \hyperlink{tab:3}{Table 3}). The CPU
walk also needs the ragged layout (\hyperlink{sec:5.2}{Section 5.2}). A padded CPU walk is
about 22 times slower in a walk microbenchmark (Figure 8) and loses the
whole saving at light load.

Hardware counters and a concurrency sweep both point to waiting as the
cost, though neither counts launches directly. Across all 36
implementations, streaming multiprocessor activity stays between 3.42
and 5.42 percent, and occupancy (share of thread slots used) never
exceeds 0.86 percent, while reported utilization runs 12.3 to 27.8
percent (\hyperlink{tab:5}{Table 5} gives the averages). Reported utilization is 5.2 times
the multiprocessor-active share when the walk runs on the GPU and 2.9
times when it runs on the CPU.

The concurrency sweep reran the CPU trie against the control with a cap
on how many rows one decode step may hold (Appendix L). With 1 row per
step there is one group, and the CPU trie saves nothing. With 4 rows it
saves 7.7 percent, and with 16, 64 or 256 rows it saves 24 to 25
percent. The saving appears only once a step holds several groups, as a
cost paid per group predicts.

\begin{table*}[!tbp]
\centering
{\footnotesize \hypertarget{tab:6}{}\textbf{Table 6.} The fastest GPU and CPU implementations on the
item-first and store-first layouts. The control row is milliseconds per
token, and other cells are percent change against that layout's control.
The store columns repeat \hyperlink{tab:4}{Table 4}'s pooled values. The full CPU stack is
unpaired on the item layout and composed on the store one (Appendix K).
Bold marks the best in each column.\par}
\vspace{0.7ex}
\footnotesize
\begin{adjustbox}{max width=\linewidth}
\begin{tabular}{@{}l r r r r@{}}
\toprule
\thead{implementation} & \thead{item, light} & \thead{item, saturated} & \thead{store, light} & \thead{store, saturated} \\
\midrule
control & 7.8 {[}7.6, 7.9{]} & 17.9 {[}16.8, 19.0{]} & 10.6 {[}10.1,
11.1{]} & 199.0 {[}188.5, 209.6{]} \\
full GPU stack & −6.3\% {[}−9.7, −2.9{]} & −5.7\% {[}−14.4, +3.0{]} &
−21.6\% & −56.8\% \\
full CPU stack & \textbf{−51.2\% {[}−52.2, −50.1{]}} & \textbf{−75.8\%
{[}−77.3, −74.2{]}} & \textbf{−65.9\%} & \textbf{−97.9\%} \\
\bottomrule
\end{tabular}
\end{adjustbox}
\end{table*}

\begin{table*}[!tbp]
\centering
{\footnotesize \hypertarget{tab:7}{}\textbf{Table 7.} This paper beside the closest prior systems. Wide
beam: many beams over one short identifier. Bold marks our column.\par}
\vspace{0.7ex}
\footnotesize
\begin{adjustbox}{max width=\linewidth}
\begin{tabular}{@{}l l l l l l@{}}
\toprule
\thead{} & \thead{FlashTrie} & \thead{STATIC} & \thead{GRACE} & \thead{XGrammar} & \thead{this paper} \\
\midrule
fixed ahead of time & yes & yes, per corpus & no, per request & yes &
\textbf{no, per token} \\
changes mid sequence & no & no & no & no & \textbf{yes, blocked set
grows} \\
identifiers per request & 1 & 1 & many & n/a & \textbf{about 110 per
page} \\
structures walked & 1 trie & 1 trie & 1 trie + filter & 1 grammar &
\textbf{2 tries + 3 indexes} \\
walk placement & GPU & GPU & GPU & CPU & \textbf{both, compared} \\
decoding & wide beam, short & wide beam, short & wide beam, short & n/a
& \textbf{three candidates, 843 tokens} \\
\bottomrule
\end{tabular}
\end{adjustbox}
\end{table*}

\hyperlink{alg:4}{Algorithm 4} queues no GPU work until every walk is done, so the walks
run under the forward pass, and the CPU knows an empty list without a
read-back. Against the control, the CPU recipe is 1.25 times as fast.
Adding the overlap raises this to 1.53 times, and adding the numpy trie
and the batched write raises it to 2.93 times, so the three additions
multiply speed by 1.25, 1.22 and 1.92. A constrained token then costs
about 1.7 times an unconstrained one (2.09 ms, \hyperlink{tab:1}{Table 1}), against about 5
times for the control.

We expect a GPU walk to win when the walk itself does substantial
arithmetic, for example a wide beam over a wide, shallow trie (up to
1,000 candidates per group, \hyperlink{sec:7}{Section 7}), and on servers with fewer than
our four CPU cores per GPU. We did not test these cases.

\subsection[5.6 A second page layout]{\hypertarget{sec:5.6}{}5.6 A second page layout}\label{a-second-page-layout}

The same catalog serves an item-first page. Its outer slot is an item
and its inner slot is the store that sells it. It keeps all three
properties of \hyperlink{sec:2.4}{Section 2.4}. Each inner carousel holds about one store,
against five items on the store-first layout, and pages are about a
third as long (281 output tokens against 843). Per request, the
constraint inverts the slice of the store-to-item index over the stores
the request can reach, so no new offline artifact is needed. \hyperlink{tab:6}{Table 6}
compares the full GPU and full CPU stacks, the fastest of each placement
on the store-first layout, on this layout.

The item-first control slows only 2.3 times from light to saturated
load, against about 19 times on the store-first layout. The full CPU
stack is fastest here too. The full GPU stack keeps under a third of its
store-first gain, at light load only.

\subsection[5.7 Incremental state and page
length]{\hypertarget{sec:5.7}{}5.7 Incremental state and page
length}\label{incremental-state-and-page-length}

Incremental state carries the trie position from step to step. Without
it, the decoder replays the prefix from the root at every step, which is
53.9 percent slower at light load and 91.1 percent slower when
saturated. The replay penalty grows from −1.2 percent at 5 store slots
to +91.8 percent at 40, more slowly than page length (Table 16).

\section[6. Lessons and limitations]{\hypertarget{sec:6}{}6. Lessons and limitations}\label{lessons-and-limitations}

We suggest this order to a team adding a LEGIT constraint.

\begin{enumerate}
\def\labelenumi{\arabic{enumi}.}
\tightlist
\item
  Keep the trie and build the structures first (\hyperlink{sec:4}{Section 4}): compressed
  sparse row tries, depth-first leaf numbers, every filter a sorted
  array, and the blocked count in the legal-child test.
\item
  Walk on the CPU, over a ragged trie (\hyperlink{sec:5.2}{Section 5.2}), then add the two
  changes of the CPU recipe, inner subtrie compaction and a cached
  blocked set per row. Neither needs a custom kernel.
\item
  Take the walk off the critical path (\hyperlink{sec:5.5}{Section 5.5}): walk every group
  under the forward pass, answer trie queries with plain arrays, and
  write every list in one copy.
\item
  Carry the trie position from step to step (\hyperlink{sec:5.7}{Section 5.7}). Replaying the
  prefix costs more as pages grow.
\item
  Measure each change where it ships, in both load regimes (\hyperlink{sec:5.1}{Section
  5.1}). Effects do not add, and a change can flip sign with placement,
  layout or load.
\end{enumerate}

The evidence has four limits.

\begin{itemize}
\tightlist
\item
  We measure speed and page validity but not recommendation quality.
\item
  We have no like-for-like external baseline. The grammar engine we
  tried was not both exact and fast, and the GPU trie systems target
  wide beams over short identifiers, a different regime (\hyperlink{sec:7}{Section 7}).
\item
  We test one system, one domain and one hardware shape: a 0.6 billion
  parameter model on pages of twenty stores, twice production's ten.
  Shorter pages would shrink every percentage. We used one B200 GPU and
  four CPU cores per run. Fewer or slower cores may shrink the CPU
  walk's lead (Appendix G).
\item
  Several entries can share one Semantic ID. Stores outnumber their
  Semantic IDs by 5.7 percent, and about ten items share each item ID.
  The constraint allows an ID if any entry sharing it is eligible, so
  mapping a generated ID back to an entry must choose an eligible one.
  Our catalog check covers neither this step nor the one-store-per-chain
  rule (Appendix C).
\end{itemize}

\section[7. Related work]{\hypertarget{sec:7}{}7. Related work}\label{related-work}

\hyperlink{tab:7}{Table 7} places LEGIT beside the closest prior systems. Each enforces a
constraint fixed ahead of time or per request, and each walks a single
trie or grammar.

\textbf{From generative retrieval to generative recommenders.} TIGER
brought Semantic IDs to recommendation, and trie-constrained decoding is
the standard way to keep generated identifiers real. Generative
recommenders now write whole pages and slates (\hyperlink{sec:1}{Section 1}). GenPage
enforces business rules with a per-token mask over one-token entities.
None of the three reports the constraint's cost.

\textbf{Grammar-constrained decoding.} XGrammar, Outlines and llguidance
compile a grammar ahead of time and apply it as a token mask, and SGLang
\citep{zheng2024sglang} adds forced runs of tokens. XGrammar precomputes
masks for context-independent tokens, but almost none of a LEGIT legal
set is context independent, since it depends on the map cell and the
page so far. We tried three encodings of this constraint for a grammar
engine (Appendix M): the exact one does not compile, a factored one
accepts all 6 × 10¹⁷ item tuples, and a CPU interpreter is 100 times
slower. vLLM's per-request allowed-token list is fixed for the request;
a LEGIT legal set changes at every token.

\textbf{Accelerating the trie walk.} FlashTrie walks a bit-compressed
trie on the GPU, up to 24 times faster than a tuned CPU baseline at beam
widths up to 1,000. STATIC compiles a business-logic subset into one
sparse matrix per corpus and reports 948 times over a CPU trie. GRACE,
the closest work, enforces advertiser eligibility per request in a wide
beam over short sequences. All three put the walk on the GPU, where a
wide, shallow walk's arithmetic may dominate. Our walk is narrow and
deep, with three candidates over pages of 843 tokens.

\section[8. Conclusion]{\hypertarget{sec:8}{}8. Conclusion}\label{conclusion}

We built a LEGIT constraint for pages of stores and items and
benchmarked 36 ways to enforce it on one B200 GPU. Without a trie,
output names entries that do not exist. With it, every page passes our
independent catalog check that every entry exists, is eligible and
appears once. One trie per kind of entry, an index between them and one
shared leaf numbering suffice, and the same offline artifacts served a
second layout. Our evidence points to kernel launches and waits as the
cost. The full CPU stack, which walks the trie on the CPU while the GPU
runs the model and needs no custom kernel, cuts time per output token by
65.9 percent against the control, to about 1.7 times the unconstrained
cost.

\section{References}\label{references}

\renewcommand{\bibsection}{}
\providecommand{\doi}[1]{doi:\href{https://doi.org/#1}{#1}}
\bibliographystyle{mlsys2025}
\bibliography{paper_short}

\begin{thebibliography}{22}
\providecommand{\natexlab}[1]{#1}
\providecommand{\url}[1]{\texttt{#1}}
\expandafter\ifx\csname urlstyle\endcsname\relax
  \providecommand{\doi}[1]{doi: #1}\else
  \providecommand{\doi}{doi: \begingroup \urlstyle{rm}\Url}\fi

\bibitem[Anandakumar et~al.(2026)Anandakumar, Mukkara, Hu, Chen, Kumar, Ye,
  Lou, and Jiao]{anandakumar2026flashtrie}
Anandakumar, D., Mukkara, A., Hu, W., Chen, J., Kumar, M.~A., Ye, T., Lou, Q.,
  and Jiao, J.
\newblock {FlashTrie: A GPU-Accelerated Constrained Beam Search for Generative
  Retrieval}.
\newblock \emph{arXiv preprint arXiv:2607.10044}, 2026.

\bibitem[Brodsky(2018)]{brodsky2018h3}
Brodsky, I.
\newblock {H3: Uber's Hexagonal Hierarchical Spatial Index}.
\newblock Uber Engineering blog, 2018.
\newblock URL \url{https://www.uber.com/blog/h3/}.

\bibitem[D'Amico et~al.(2026)D'Amico, {De Nadai}, Chandar, Vohra, Lin, Lefarov,
  Gigioli, Penha, Kopysitsky, Senese, Mei, Fabbri, Semerci, Zhao, Tang, {St.
  Thomas}, Ranieri, Smith, Bernkopf, Leung, Fazelnia, VanMiddlesworth, Heath,
  Skiden, Wang, Cole, Damianou, Hristakeva, Wilbur, Chillara, Radosavljevic,
  Chitkara, Adapa, Elenter, Huber, Wood, Vedantam, Stypka, Ghael, Gould,
  Murgatroyd, Raimond, Lalmas, and Bennett]{damico2026glide}
D'Amico, E., {De Nadai}, M., Chandar, P., Vohra, D., Lin, S., Lefarov, M.,
  Gigioli, P., Penha, G., Kopysitsky, I., Senese, I.~J., Mei, D., Fabbri, F.,
  Semerci, O., Zhao, Y., Tang, V., {St. Thomas}, B., Ranieri, A., Smith, M.
  N.~K., Bernkopf, A., Leung, B., Fazelnia, G., VanMiddlesworth, M., Heath,
  T.~C., Skiden, P.~P., Wang, A.~Y., Cole, D.~J., Damianou, A., Hristakeva, M.,
  Wilbur, R., Chillara, T., Radosavljevic, V., Chitkara, P., Adapa, S.,
  Elenter, J., Huber, B., Wood, J., Vedantam, S., Stypka, J., Ghael, S., Gould,
  M.~D., Murgatroyd, D., Raimond, Y., Lalmas, M., and Bennett, P.~N.
\newblock {Deploying Semantic ID-based Generative Retrieval for Large-Scale
  Podcast Discovery at Spotify}.
\newblock \emph{arXiv preprint arXiv:2603.17540}, 2026.

\bibitem[{De Cao} et~al.(2021){De Cao}, Izacard, Riedel, and
  Petroni]{decao2021genre}
{De Cao}, N., Izacard, G., Riedel, S., and Petroni, F.
\newblock {Autoregressive Entity Retrieval}.
\newblock In \emph{International Conference on Learning Representations}, 2021.

\bibitem[Dong et~al.(2025)Dong, Ruan, Cai, Lai, Xu, Zhao, and
  Chen]{dong2025xgrammar}
Dong, Y., Ruan, C.~F., Cai, Y., Lai, R., Xu, Z., Zhao, Y., and Chen, T.
\newblock {XGrammar: Flexible and Efficient Structured Generation Engine for
  Large Language Models}.
\newblock In \emph{Proceedings of Machine Learning and Systems}, volume~7,
  2025.

\bibitem[Fang et~al.(2026)Fang, Huang, Zhang, He, Xiao, Li, Yetim, Yang, Wei,
  Tian, Wang, Sun, Li, Yan, and Liu]{fang2026grace}
Fang, Z., Huang, Y., Zhang, A., He, Y., Xiao, R., Li, C., Yetim, Y., Yang, S.,
  Wei, X., Tian, F., Wang, L., Sun, C., Li, L., Yan, N., and Liu, G.
\newblock {GRACE: Generative Recommender Acceleration Engine for Real-Time Ads
  Retrieval}.
\newblock \emph{arXiv preprint arXiv:2608.00938}, 2026.

\bibitem[Ge et~al.(2014)Ge, He, Ke, and Sun]{ge2014opq}
Ge, T., He, K., Ke, Q., and Sun, J.
\newblock {Optimized Product Quantization}.
\newblock \emph{IEEE Transactions on Pattern Analysis and Machine
  Intelligence}, 2014.

\bibitem[Geng et~al.(2025)Geng, Cooper, Moskal, Jenkins, Berman, Ranchin, West,
  Horvitz, and Nori]{geng2025jsonschemabench}
Geng, S., Cooper, H., Moskal, M., Jenkins, S., Berman, J., Ranchin, N., West,
  R., Horvitz, E., and Nori, H.
\newblock {JSONSchemaBench: A Rigorous Benchmark of Structured Outputs for
  Language Models}.
\newblock \emph{arXiv preprint arXiv:2501.10868}, 2025.
\newblock Describes and evaluates llguidance, software at
  guidance-ai/llguidance.

\bibitem[He et~al.(2026)He, Heldt, Hong, Keshavan, Mao, Mehta, Su, Tsai, Wang,
  Wang, Yi, Baugher, Cakici, Chi, Goodrow, Han, Ma, Rosales, {Van Soest},
  Tandon, Wu, Yang, and Zheng]{he2026plum}
He, R., Heldt, L., Hong, L., Keshavan, R., Mao, S., Mehta, N., Su, Z., Tsai,
  A., Wang, Y., Wang, S.-C., Yi, X., Baugher, L., Cakici, B., Chi, E., Goodrow,
  C., Han, N., Ma, H., Rosales, R., {Van Soest}, A., Tandon, D., Wu, S.-L.,
  Yang, W., and Zheng, Y.
\newblock {PLUM: Adapting Pre-trained Language Models for Industrial-scale
  Generative Recommendations}.
\newblock In \emph{The ACM Web Conference}, 2026.

\bibitem[Hu et~al.(2026)Hu, Guo, Ma, Li, Qin, Li, Deng, Chen, and
  Zhan]{hu2026once}
Hu, Y., Guo, J., Ma, J., Li, N., Qin, J., Li, Y., Deng, Y., Chen, X., and Zhan,
  K.
\newblock {Once Generated, Ranked: End-to-End Generative Slate Recommendation
  with Unified Semantic-Collaborative IDs}.
\newblock \emph{arXiv preprint arXiv:2608.17613}, 2026.

\bibitem[Kwon et~al.(2023)Kwon, Li, Zhuang, Sheng, Zheng, Yu, Gonzalez, Zhang,
  and Stoica]{kwon2023vllm}
Kwon, W., Li, Z., Zhuang, S., Sheng, Y., Zheng, L., Yu, C.~H., Gonzalez, J.~E.,
  Zhang, H., and Stoica, I.
\newblock {Efficient Memory Management for Large Language Model Serving with
  PagedAttention}.
\newblock In \emph{Symposium on Operating Systems Principles}, 2023.
\newblock Software, vllm-project/vllm.

\bibitem[Lee et~al.(2022)Lee, Kim, Kim, Cho, and Han]{lee2022rq}
Lee, D., Kim, C., Kim, S., Cho, M., and Han, W.-S.
\newblock {Autoregressive Image Generation using Residual Quantization}.
\newblock In \emph{Conference on Computer Vision and Pattern Recognition},
  2022.

\bibitem[{OneRec Team}(2026)]{onerec2026onereason}
{OneRec Team}.
\newblock {OneReason Technical Report}.
\newblock \emph{arXiv preprint arXiv:2606.06260}, 2026.

\bibitem[Pang et~al.(2025)Pang, Liu, Li, Zhu, Luo, Yu, Wu, Shen, Xu, Wang,
  Jiang, Zhuo, and Li]{pang2025higr}
Pang, Y., Liu, Z., Li, Y., Zhu, S., Luo, Z., Yu, C., Wu, S., Shen, S., Xu, C.,
  Wang, B., Jiang, K., Zhuo, C., and Li, Z.
\newblock {HiGR: Industrial-Scale Hierarchical Generative Slate Recommendation
  Framework in Tencent}.
\newblock \emph{arXiv preprint arXiv:2512.24787}, 2025.

\bibitem[Rajput et~al.(2023)Rajput, Mehta, Singh, Keshavan, Vu, Heldt, Hong,
  Tay, Tran, Samost, Kula, Chi, and Sathiamoorthy]{rajput2023tiger}
Rajput, S., Mehta, N., Singh, A., Keshavan, R.~H., Vu, T., Heldt, L., Hong, L.,
  Tay, Y., Tran, V.~Q., Samost, J., Kula, M., Chi, E.~H., and Sathiamoorthy, M.
\newblock {Recommender Systems with Generative Retrieval}.
\newblock In \emph{Advances in Neural Information Processing Systems},
  volume~36, 2023.

\bibitem[Su et~al.(2026)Su, Katsman, Wang, He, Heldt, Keshavan, Wang, Yi, Gao,
  Dalal, Hong, Chi, and Han]{su2026static}
Su, Z., Katsman, I., Wang, Y., He, R., Heldt, L., Keshavan, R., Wang, S.-C.,
  Yi, X., Gao, M., Dalal, O., Hong, L., Chi, E., and Han, N.
\newblock {Vectorizing the Trie: Efficient Constrained Decoding for LLM-based
  Generative Retrieval on Accelerators}.
\newblock In \emph{ACM SIGKDD Conference on Knowledge Discovery and Data
  Mining}, 2026.

\bibitem[Tay et~al.(2022)Tay, Tran, Dehghani, Ni, Bahri, Mehta, Qin, Hui, Zhao,
  Gupta, Schuster, Cohen, and Metzler]{tay2022dsi}
Tay, Y., Tran, V.~Q., Dehghani, M., Ni, J., Bahri, D., Mehta, H., Qin, Z., Hui,
  K., Zhao, Z., Gupta, J., Schuster, T., Cohen, W.~W., and Metzler, D.
\newblock {Transformer Memory as a Differentiable Search Index}.
\newblock In \emph{Advances in Neural Information Processing Systems},
  volume~35, 2022.

\bibitem[Wang et~al.(2026)Wang, Pan, and Baltrunas]{wang2026genpage}
Wang, L., Pan, J., and Baltrunas, L.
\newblock {GenPage: Towards End-to-End Generative Homepage Construction at
  Netflix}.
\newblock In \emph{Proceedings of the 20th ACM Conference on Recommender
  Systems}, 2026.
\newblock \doi{10.1145/3773078.3831915}.

\bibitem[Wang et~al.(2022)Wang, Hou, Wang, Miao, Wu, Sun, Chen, Xia, Chi, Zhao,
  Liu, Xie, Sun, Deng, Zhang, and Yang]{wang2022nci}
Wang, Y., Hou, Y., Wang, H., Miao, Z., Wu, S., Sun, H., Chen, Q., Xia, Y., Chi,
  C., Zhao, G., Liu, Z., Xie, X., Sun, H.~A., Deng, W., Zhang, Q., and Yang, M.
\newblock {A Neural Corpus Indexer for Document Retrieval}.
\newblock In \emph{Advances in Neural Information Processing Systems},
  volume~35, 2022.

\bibitem[Willard \& Louf(2023)Willard and Louf]{willard2023outlines}
Willard, B.~T. and Louf, R.
\newblock {Efficient Guided Generation for Large Language Models}.
\newblock \emph{arXiv preprint arXiv:2307.09702}, 2023.

\bibitem[Yang et~al.(2025)Yang, Li, Yang, Zhang, Hui, Zheng, Yu, Gao, Huang,
  Lv, Zheng, Liu, Zhou, Huang, Hu, Ge, Wei, Lin, Tang, Yang, Tu, Zhang, Yang,
  Yang, Zhou, Zhou, Lin, Dang, Bao, Yang, Yu, Deng, Li, Xue, Li, Zhang, Wang,
  Zhu, Men, Gao, Liu, Luo, Li, Tang, Yin, Ren, Wang, Zhang, Ren, Fan, Su,
  Zhang, Zhang, Wan, Liu, Wang, Cui, Zhang, Zhou, and Qiu]{yang2025qwen3}
Yang, A., Li, A., Yang, B., Zhang, B., Hui, B., Zheng, B., Yu, B., Gao, C.,
  Huang, C., Lv, C., Zheng, C., Liu, D., Zhou, F., Huang, F., Hu, F., Ge, H.,
  Wei, H., Lin, H., Tang, J., Yang, J., Tu, J., Zhang, J., Yang, J., Yang, J.,
  Zhou, J., Zhou, J., Lin, J., Dang, K., Bao, K., Yang, K., Yu, L., Deng, L.,
  Li, M., Xue, M., Li, M., Zhang, P., Wang, P., Zhu, Q., Men, R., Gao, R., Liu,
  S., Luo, S., Li, T., Tang, T., Yin, W., Ren, X., Wang, X., Zhang, X., Ren,
  X., Fan, Y., Su, Y., Zhang, Y., Zhang, Y., Wan, Y., Liu, Y., Wang, Z., Cui,
  Z., Zhang, Z., Zhou, Z., and Qiu, Z.
\newblock {Qwen3 Technical Report}.
\newblock \emph{arXiv preprint arXiv:2505.09388}, 2025.

\bibitem[Zheng et~al.(2024)Zheng, Yin, Xie, Sun, Huang, Yu, Cao, Kozyrakis,
  Stoica, Gonzalez, Barrett, and Sheng]{zheng2024sglang}
Zheng, L., Yin, L., Xie, Z., Sun, C., Huang, J., Yu, C.~H., Cao, S., Kozyrakis,
  C., Stoica, I., Gonzalez, J.~E., Barrett, C., and Sheng, Y.
\newblock {SGLang: Efficient Execution of Structured Language Model Programs}.
\newblock In \emph{Advances in Neural Information Processing Systems},
  volume~37, 2024.

\end{thebibliography}

\section{AI Usage Disclosure}\label{ai-usage-disclosure}

We used Claude Opus models, through Claude Code, in this work.
\textbf{Code:} mostly AI. The models wrote most of the constraint
implementations, the benchmark harness, the analysis scripts and the
figure generators, under the authors' direction and review.
\textbf{Writing:} part AI, part human. The models drafted and revised
text that the authors directed and edited. \textbf{Literature search:}
part AI, part human. The models helped find and summarize prior work;
the authors chose what to cite. \textbf{Experiments and data:} human.
The authors launched every benchmark sweep, ran the statistics with the
model-written scripts, and built the catalog tables. \textbf{Human
only:} research ideation, experiment design and the review of drafts.
The authors take full responsibility for the paper's content, including
every citation.

\clearpage
\onecolumn
\section{Appendix}\label{appendix}

\subsection[A. How the counts add up]{\hypertarget{sec:A}{}A. How the counts add up}\label{a.-how-the-counts-add-up}

Terms used throughout this appendix: the CPU recipe is the CPU trie with
inner subtrie compaction and the memo; the overlap, the numpy trie and
the batched write are its three extensions (\hyperlink{sec:K}{Appendix K}), and the full
CPU stack is the recipe with all three.

The benchmark ran as nine sweeps, with 100 requests per run. Each sweep
ran every one of its implementations in every round and both regimes,
the control included in all but sweep 9.

\begin{table}[H]
\centering
{\small \hypertarget{tab:8}{}\textbf{Table 8.} The nine sweeps. Implementations include the control,
which sweep 9 did not run; a run is one implementation in one round and
one regime, with 100 requests.\par}
\vspace{0.8ex}
\footnotesize
\begin{adjustbox}{max width=\linewidth}
\begin{tabular}{@{}l l r r r r@{}}
\toprule
\thead{sweep} & \thead{dates} & \thead{implementations} & \thead{rounds per regime} & \thead{runs} & \thead{pages} \\
\midrule
1, the matrix & September 23 to 24 & 22 & 6 & 262 & 26,200 \\
2, CPU trie variants & September 24 & 4 & 6 & 48 & 4,800 \\
3, compositions & September 24 & 10 & 6 & 120 & 12,000 \\
4, leaders replicated & September 24 & 5 & 6 & 60 & 6,000 \\
5, the matrix replicated & September 25 & 29 & 6 & 348 & 34,800 \\
6, leveling & September 25 & 9 & 6 & 108 & 10,800 \\
7, CPU recipe, with and without memo & September 27 & 3 & 12 & 72 &
7,200 \\
8, CPU recipe and overlap, two models & September 30 & 3 & 12 & 144 &
14,400 \\
9, numpy trie and batched write, two models & September 30 to October 1
& 2 & 12 & 96 & 9,600 \\
\bottomrule
\end{tabular}
\end{adjustbox}
\end{table}

\begin{itemize}
\tightlist
\item
  \textbf{36 implementations} are 35 alternatives and the control.
  Implementations recur across sweeps, so the per sweep counts sum to
  more than 36. Sweeps 8 and 9 each ran two models, the 0.6 billion
  parameter one and a 4 billion parameter one. Both count in the runs;
  only the first enters the ranking. \hyperlink{sec:K}{Appendix K} reports the larger
  model's sweep 8 numbers; its sweep 9 runs are page checked but not
  analysed.
\item
  \textbf{12 rounds per regime.} An implementation in several sweeps
  enters the ranking with its latest 12 rounds per regime, pooled across
  sweeps. The control is in every sweep but the last and carries 60.
\item
  \textbf{70 combinations} are the 35 alternatives in two regimes, one
  ranking test each.
\item
  \textbf{Page checks.} The shape and catalog checks cover every page of
  all nine sweeps (\hyperlink{sec:C}{Appendix C}). The content comparison covers 64
  combinations in sweeps 1 to 6, the implementation added in sweep 7
  against that sweep's floor, and sweeps 8 and 9 against their own
  baselines.
\item
  \textbf{Control values.} Each effect uses the control of its own
  rounds, so the control's absolute time differs by source. Light, in ms
  per token: 10.22 in the six rounds of Table 1, \hyperlink{tab:10}{10}.6 in sweep 6 (Table
  6), 10.30 in sweep 8 (\hyperlink{tab:17}{Table 17}), and 10.56 pooled over the rounds the
  ranking uses (Tables 4 and \hyperlink{tab:9}{9}). Saturated: 186.31, 199.0 and 188.0 for
  the first three. Its throughput is 0.343 pages per second over its 48
  saturated runs in sweeps 1 to 7 and 0.350 in sweep 8.
\item
  \textbf{1,258 runs} rather than 1,260: sweep 1 is missing two runs.
\item
  \textbf{Outside these counts} are the side experiments: on the second
  page layout (Section 5.6), a confirmation sweep of five
  implementations and a later one of the full CPU stack, each twelve
  rounds per regime; and the page length sweep (\hyperlink{sec:I}{Appendix I}).
\end{itemize}

\subsection[B. The full ranking]{\hypertarget{sec:B}{}B. The full ranking}\label{b.-the-full-ranking}

\begin{table}[H]
\centering
{\small \hypertarget{tab:9}{}\textbf{Table 9.} The ranking, all 35 implementations against the
control, grouped by where the walk executes, simplest first. \hyperlink{sec:J}{Appendix J}
draws the implementations. Lower is faster. Brackets are 95 percent
intervals, and n.s. marks one spanning zero. The full CPU stack is
composed through the overlap row, because its sweep ran no control.
Every implementation is on its latest twelve rounds per regime beside a
control. Tokens per second is for one request at light load, implied by
the light effect and the control row's pooled time. Bold marks the best
point estimate in each group, not a significant lead. The fused stack is
the fused kernel with the batched launch and the deferred check.\par}
\vspace{0.8ex}
\footnotesize
\begin{adjustbox}{max width=\linewidth}
\begin{tabular}{@{}l l l r@{}}
\toprule
\thead{implementation} & \thead{light} & \thead{saturated} & \thead{tokens/s, light} \\
\midrule
control: padded trie on the GPU & 10.56 ms, the reference & 194.43 ms,
the reference & 95 \\
\emph{walk on the GPU, one change} & & & \\
memo & −1.9\% {[}−5.6, +1.8{]}, n.s. & \textbf{−10.4\% {[}−16.6,
−4.2{]}} & 97 \\
candidate cache & −0.4\% {[}−4.1, +3.3{]}, n.s. & +7.3\% {[}+0.9,
+13.6{]} & 95 \\
blocked bitset & +2.3\% {[}−1.6, +6.1{]}, n.s. & +6.5\% {[}+0.5,
+12.6{]} & 93 \\
deferred check & \textbf{−6.2\% {[}−8.8, −3.6{]}} & −6.3\% {[}−12.1,
−0.6{]} & \textbf{101} \\
inner subtrie compaction & +2.2\% {[}−1.1, +5.6{]}, n.s. & +9.1\%
{[}+3.0, +15.3{]} & 93 \\
outer subtrie compaction & +0.3\% {[}−3.6, +4.2{]}, n.s. & +0.5\%
{[}−5.7, +6.7{]}, n.s. & 94 \\
ragged children on the GPU & +113.2\% {[}+101.8, +124.7{]} & +106.4\%
{[}+99.0, +113.8{]} & 44 \\
prefetch on a side stream & +14.1\% {[}+10.6, +17.5{]} & +4.6\% {[}+0.3,
+9.0{]} & 83 \\
batched chain exclusion & +35.4\% {[}+27.2, +43.5{]} & +105.5\%
{[}+90.5, +120.4{]} & 70 \\
batched chain exclusion + memo & +32.2\% {[}+26.6, +37.9{]} & +110.8\%
{[}+99.1, +122.5{]} & 72 \\
\emph{walk on the GPU, fused kernels} & & & \\
fused kernel & −8.9\% {[}−11.9, −6.0{]} & −11.0\% {[}−17.1, −4.9{]} &
104 \\
fused kernel + kernel profiling & −2.5\% {[}−5.6, +0.7{]}, n.s. & +1.8\%
{[}−3.0, +6.7{]}, n.s. & 97 \\
fused mask apply & −10.1\% {[}−12.8, −7.4{]} & −12.9\% {[}−18.7, −7.2{]}
& 105 \\
fused kernel + batched launch & −19.4\% {[}−22.0, −16.8{]} & −42.2\%
{[}−47.5, −36.9{]} & 117 \\
fused kernel + batched launch + pinned mask index & −18.5\% {[}−21.3,
−15.7{]} & −43.5\% {[}−48.2, −38.7{]} & 116 \\
fused stack & −16.8\% {[}−20.1, −13.6{]} & −49.8\% {[}−53.9, −45.6{]} &
114 \\
full GPU stack & \textbf{−21.6\% {[}−24.8, −18.3{]}} & \textbf{−56.8\%
{[}−61.1, −52.4{]}} & \textbf{121} \\
fused stack + candidate cache & −19.7\% {[}−23.1, −16.4{]} & −45.2\%
{[}−51.9, −38.5{]} & 118 \\
fused stack + graph mask step & −19.8\% {[}−22.9, −16.7{]} & −47.4\%
{[}−51.4, −43.4{]} & 118 \\
fused stack + graph mask step + pinned mask index & −19.1\% {[}−22.7,
−15.5{]} & −43.3\% {[}−48.6, −38.0{]} & 117 \\
fused stack + fused mask apply & −12.0\% {[}−15.7, −8.4{]} & −21.3\%
{[}−26.7, −15.9{]} & 108 \\
\emph{walk on the CPU} & & & \\
CPU trie & −10.0\% {[}−13.3, −6.6{]} & −41.2\% {[}−45.2, −37.2{]} &
105 \\
CPU trie + padded layout & +3.9\% {[}+0.5, +7.2{]} & −11.6\% {[}−15.7,
−7.5{]} & 91 \\
CPU trie + memo & −16.2\% {[}−19.5, −12.9{]} & −43.8\% {[}−48.0,
−39.5{]} & 113 \\
CPU trie + candidate cache & −11.0\% {[}−14.9, −7.0{]} & −37.4\%
{[}−43.1, −31.8{]} & 106 \\
CPU trie + deferred check & −13.3\% {[}−16.6, −9.9{]} & −37.1\%
{[}−41.0, −33.2{]} & 109 \\
CPU trie + deferred check + memo & −14.9\% {[}−18.2, −11.5{]} & −42.0\%
{[}−46.8, −37.1{]} & 111 \\
CPU trie + outer subtrie compaction & −13.2\% {[}−17.1, −9.3{]} &
−40.1\% {[}−44.9, −35.3{]} & 109 \\
CPU trie + outer subtrie compaction + blocked bitset & −14.2\% {[}−17.7,
−10.6{]} & −38.4\% {[}−43.2, −33.7{]} & 110 \\
CPU trie + inner subtrie compaction & −16.7\% {[}−20.3, −13.1{]} &
−55.1\% {[}−60.3, −49.9{]} & 114 \\
CPU recipe & −20.2\% {[}−22.5, −17.8{]} & −63.5\% {[}−69.0, −58.0{]} &
119 \\
CPU recipe + overlap & −34.7\% {[}−37.0, −32.3{]} & −77.6\% {[}−82.9,
−72.3{]} & 145 \\
full CPU stack & \textbf{−65.9\% {[}−67.5, −64.3{]}} & \textbf{−97.9\%
{[}−98.2, −97.5{]}} & \textbf{278} \\
\emph{ablations} & & & \\
no row grouping & +2.1\% {[}−2.1, +6.4{]}, n.s. & +7.8\% {[}+1.2,
+14.4{]} & 93 \\
replay the prefix each step & +53.9\% {[}+49.7, +58.0{]} & +91.1\%
{[}+83.8, +98.4{]} & 62 \\
\bottomrule
\end{tabular}
\end{adjustbox}
\end{table}

\begin{itemize}
\tightlist
\item
  \textbf{Multiple tests.} Each ranking test is a Welch test, because
  variance differs between implementations. 61 of the 70 ranking tests
  reach p below 0.05, 53 survive a Bonferroni threshold over 70 tests,
  and a Benjamini-Hochberg procedure at a 5 percent false discovery rate
  keeps all 61. No claim we rely on is among those dropped.
\end{itemize}

\subsection[C. Page validity]{\hypertarget{sec:C}{}C. Page validity}\label{c.-page-validity}

Two checks cover validity. The first compares implementations with each
other: every implementation must build pages of the same shape (\hyperlink{tab:10}{Table
10}) from the same stores.

\begin{itemize}
\tightlist
\item
  \textbf{Same pages.} Over every page of the 0.6 billion parameter
  model in all nine sweeps, 116,100 pages in 176 cells of sweep,
  implementation and regime (113,800 from the runs of \hyperlink{tab:8}{Table 8} and 2,300
  from sweep 9's numpy trie alone arm, which \hyperlink{tab:8}{Table 8} does not count),
  every implementation builds 19.09 stores per page and the same 843
  token median. The 4 billion parameter model builds 93.25 items and
  0.76 underfilled slots per page, the same across its implementations,
  so the model sets that shape. We sample pages at three candidates, so
  two runs of the same code differ and a page hash has no power. We
  compare instead the store overlap with a baseline against the
  baseline's overlap with itself on another round, 0.1615 for the
  control at light load. All 64 combinations of sweeps 1 to 6 sit on
  that floor, as do the implementation added in sweep 7 and all 16
  combinations of sweeps 8 and 9, each within 0.009 of its own sweep's
  floor. No implementation changes its stores.
\end{itemize}

\begin{table}[H]
\centering
{\small \hypertarget{tab:10}{}\textbf{Table 10.} The correctness gate over every page of the 0.6
billion parameter model in all nine sweeps, 116,100 pages in 176 cells
of sweep, implementation and regime. Its parse column did not populate,
so the independent check of \hyperlink{tab:11}{Table 11} stands in for it.\par}
\vspace{0.8ex}
\footnotesize
\begin{adjustbox}{max width=\linewidth}
\begin{tabular}{@{}l l l@{}}
\toprule
\thead{invariant} & \thead{result} & \thead{spread across the 176 cells} \\
\midrule
stores per page & 19.09 & identical to two decimals in every cell \\
items per page & 94.31 & 94.05 to 94.55 \\
underfilled store slots per page & 0.39 & 0.32 to 0.46 \\
median output length & 843 tokens & identical in every cell \\
pages with zero stores & 3.000 percent & identical to three decimals in
every cell \\
\bottomrule
\end{tabular}
\end{adjustbox}
\end{table}

The second compares each page with the catalog. We parse every decoded
page and join its stores and items against the catalog tables and the
request's map cell, with SQL written apart from the decoder, so a bug
every implementation shared would still be caught. We checked all
122,026 pages with a store in the nine sweeps; the other 3,774 hold no
store, because three fixed requests return an empty page in every run.
The two counts sum to \hyperlink{tab:8}{Table 8}'s 125,800 pages, so this check covers the
runs of \hyperlink{tab:8}{Table 8}; the 2,300 numpy trie only pages of sweep 9 are covered
by the gate of \hyperlink{tab:10}{Table 10} alone. The last row is the ablation arm without
the location filter: the same join finds only 63 of its 24,000 stores in
reach, so the check can see a violation. Chain once is not checked,
because the chain table covers only part of the catalog. The item first
pages of Section 5.6 pass the same check with the nesting reversed:
296,296 items, each in the item table, and 315,047 stores under them,
each selling its item and in reach of the request, with no repeats and
no violation.

\begin{table}[H]
\centering
{\small \hypertarget{tab:11}{}\textbf{Table 11.} An independent catalog check of every decoded page of
all nine sweeps, all implementations and regimes. The last row is a
positive control: the ablation arm without the location filter.\par}
\vspace{0.8ex}
\footnotesize
\begin{adjustbox}{max width=\linewidth}
\begin{tabular}{@{}l l l@{}}
\toprule
\thead{check} & \thead{identifiers checked} & \thead{violations} \\
\midrule
store exists in the catalog & 2,401,522 stores & 0 \\
store delivers to the request's map cell & 2,401,522 stores & 0 \\
store not repeated on its page & 2,401,522 stores & 0 \\
item on the menu of a store with that identifier & 11,851,173 items &
0 \\
item not repeated in its row & 11,851,173 items & 0 \\
positive control: location filter removed & 24,000 stores & 23,937 out
of reach \\
\bottomrule
\end{tabular}
\end{adjustbox}
\end{table}

\subsection[D. Throughput and tail
latency]{\hypertarget{sec:D}{}D. Throughput and tail
latency}\label{d.-throughput-and-tail-latency}

At light load every implementation keeps up with arrivals, so pages per
second equals the arrival rate, 0.148, and only latency moves. When
saturated, pages per second is the rate at which a full queue drains.
Rows are grouped as in \hyperlink{tab:9}{Table 9}.

\begin{table}[H]
\centering
{\small \hypertarget{tab:12}{}\textbf{Table 12.} Throughput, in pages per second, and tail latency.
The CPU recipe and the overlap row are on sweep 8's rounds, against that
sweep's control (0.350 pages per second); the full CPU stack, composed
through the overlap, serves about 0.69 against the 0.7 arrival rate. The
control row is absolute, with its 95 percent interval over its 48
saturated runs in sweeps 1 to 7; every other row is percent change
against the control, on the latest twelve rounds per regime and pooled
across sweeps as in the ranking, with 95 percent intervals. Higher is
better in the throughput column and negative is faster in the two
latency columns. At light load pages per second equals the arrival rate
in every implementation. Bold marks the best point estimate in each
group.\par}
\vspace{0.8ex}
\footnotesize
\begin{adjustbox}{max width=\linewidth}
\begin{tabular}{@{}l r l l@{}}
\toprule
\thead{implementation} & \thead{throughput, saturated} & \thead{p95 latency, saturated} & \thead{p95 latency, light} \\
\midrule
control: padded trie on the GPU & 0.343 {[}0.339, 0.347{]} & 178.2 s
{[}174.6, 181.8{]} & 14.6 s {[}14.2, 15.0{]} \\
\emph{walk on the GPU} & & & \\
ragged children on the GPU & −37.0\% {[}−39.4, −34.6{]} & +98.4\%
{[}+91.5, +105.3{]} & +257.6\% {[}+232.9, +282.2{]} \\
fused kernel & +7.1\% {[}+3.6, +10.6{]} & −10.4\% {[}−16.0, −4.8{]} &
−15.0\% {[}−20.6, −9.4{]} \\
fused stack & +36.2\% {[}+33.4, +39.0{]} & −46.9\% {[}−50.6, −43.1{]} &
−24.0\% {[}−30.0, −18.0{]} \\
full GPU stack & \textbf{+43.3\% {[}+40.2, +46.3{]}} & \textbf{−52.9\%
{[}−56.7, −49.1{]}} & \textbf{−28.6\% {[}−35.0, −22.2{]}} \\
\emph{walk on the CPU} & & & \\
CPU trie & +28.1\% {[}+25.7, +30.5{]} & −38.5\% {[}−42.1, −34.9{]} &
−18.9\% {[}−27.6, −10.2{]} \\
CPU trie + inner subtrie compaction & +37.4\% {[}+34.6, +40.2{]} &
−49.3\% {[}−53.9, −44.7{]} & −29.9\% {[}−36.7, −23.1{]} \\
CPU recipe & +44.2\% {[}+39.7, +48.7{]} & −55.4\% {[}−59.1, −51.6{]} &
−31.4\% {[}−36.3, −26.4{]} \\
CPU recipe + overlap & +56.6\% {[}+51.5, +61.7{]} & −66.1\% {[}−69.2,
−63.1{]} & −42.6\% {[}−47.4, −37.7{]} \\
full CPU stack & \textbf{+96.9\% {[}+90.2, +103.8{]}} & \textbf{−97.5\%
{[}−97.8, −97.2{]}} & \textbf{−76.5\% {[}−78.5, −74.3{]}} \\
\bottomrule
\end{tabular}
\end{adjustbox}
\end{table}

\subsection[E. Load amplification]{\hypertarget{sec:E}{}E. Load amplification}\label{e.-load-amplification}

\begin{table}[H]
\centering
{\small \hypertarget{tab:13}{}\textbf{Table 13.} Load amplification by family. The light and saturated
columns are percent change against the control. The ratio of the
saturated effect to the light-load effect orders implementations by how
much per-group structure they remove. The CPU family is every CPU
implementation except the CPU trie on the padded layout. It leaves out
the two overlap rows, whose saturated cells mostly measure the control's
queue.\par}
\vspace{0.8ex}
\footnotesize
\begin{adjustbox}{max width=\linewidth}
\begin{tabular}{@{}l l l l l@{}}
\toprule
\thead{family} & \thead{light} & \thead{saturated} & \thead{ratio} & \thead{what it removes} \\
\midrule
CPU trie, all 9 ragged before the overlap & −10.0 to −20.2 & −37.1 to
−63.5 & 2.70 to 4.13 & the group loop itself \\
batched family, 7 of 8 & −16.8 to −21.6 & −42.2 to −56.8 & 2.18 to 2.96
& per group launches \\
batched family + fused apply & −12.0 & −21.3 & 1.77 & some per group
launches \\
fused mask apply alone & −10.1 & −12.9 & 1.28 & one pass per group \\
fused kernel alone & −8.9 & −11.0 & 1.23 & work within a group \\
deferred check alone & −6.2 & −6.3 & 1.02 & one CPU read per group \\
ragged on the GPU & +113.2 & +106.4 & 0.94 & nothing; adds divergence \\
prefetch on a side stream & +14.1 & +4.6 & 0.33 & nothing; adds
synchronization \\
\bottomrule
\end{tabular}
\end{adjustbox}
\end{table}

\subsection[F. Stability across
rounds]{\hypertarget{sec:F}{}F. Stability across
rounds}\label{f.-stability-across-rounds}

We split sweep 7's twelve rounds into halves and estimate each effect in
each half against the control of the same rounds. The halves agree
within their intervals, and 12 rounds give intervals about half as wide
as 6.

\begin{table}[H]
\centering
{\small \hypertarget{tab:14}{}\textbf{Table 14.} The same effects estimated on each half of sweep 7's
twelve rounds, against the control of the same rounds, with 95 percent
intervals. Lower is faster. The CPU recipe is the CPU trie with inner
subtrie compaction and the memo.\par}
\vspace{0.8ex}
\footnotesize
\begin{adjustbox}{max width=\linewidth}
\begin{tabular}{@{}l r r r@{}}
\toprule
\thead{implementation, regime} & \thead{rounds 1 to 6} & \thead{rounds 7 to 12} & \thead{all twelve} \\
\midrule
CPU recipe without memo, light & −16.2\% {[}−20.7, −11.7{]} & −16.9\%
{[}−21.0, −12.8{]} & −16.6\% {[}−19.5, −13.7{]} \\
CPU recipe, light & −16.9\% {[}−19.2, −14.6{]} & −20.5\% {[}−25.8,
−15.2{]} & −18.7\% {[}−21.6, −15.8{]} \\
CPU recipe without memo, saturated & −56.0\% {[}−61.5, −50.5{]} &
−53.7\% {[}−56.7, −50.7{]} & −54.8\% {[}−57.9, −51.7{]} \\
CPU recipe, saturated & −62.4\% {[}−64.1, −60.7{]} & −62.8\% {[}−67.3,
−58.3{]} & −62.6\% {[}−64.9, −60.3{]} \\
\bottomrule
\end{tabular}
\end{adjustbox}
\end{table}

\subsection[G. Hardware and model
size]{\hypertarget{sec:G}{}G. Hardware and model
size}\label{g.-hardware-and-model-size}

Every run had its own container with one NVIDIA B200 GPU, a hard limit
of four CPU cores and 48 gigabytes of CPU memory. That is the shape of
our serving box. The CPU implementations used 2.51 of the four cores at
light load and the GPU implementations 1.58 (Table 5). We did not
measure on fewer cores, on A100 or H100 GPUs, or on another CPU. The CPU
walk reads one node at a time and needs little memory bandwidth, so we
expect it to move with single core speed. The fused kernel's gain comes
from removing launches, which we expect to cost about the same on each
of these GPUs.

Sweeps 1 to 7 measured one model, with 0.6 billion parameters. At light
load its control spends 10.22 ms per token, about 2.1 ms of them the
model and about 8.1 ms the constraint, and the CPU recipe removes about
1.9 ms of those (18.7 percent in sweep 7). If that figure stayed fixed,
a larger model's longer forward would shrink its share.

\hyperlink{sec:K}{Appendix K} measures a 4 billion parameter model on the same tries and
indexes, and the figure grows instead. At light load the CPU recipe is
28.9 percent faster on the larger model against 20.1 percent on the
smaller in the same rounds, saving 4.0 ms per token against 2.1 ms. When
saturated it is 63.6 and 63.3 percent faster. A hypothesis we did not
test: a slower model keeps more requests in flight at the same arrival
rate, so a step holds more groups, and the recipe's speedup is per
group. If so, the speedup depends on the forward's length, and the
budget of \hyperlink{sec:K}{Appendix K} holds only for the smaller model. This is one more
size and three implementations: the CPU recipe holds on a model about
seven times larger, and the full ranking is untested there.

\subsection[H. The constraint in other
applications]{\hypertarget{sec:H}{}H. The constraint in other
applications}\label{h.-the-constraint-in-other-applications}

\begin{table}[H]
\centering
{\small \hypertarget{tab:15}{}\textbf{Table 15.} The page constraint in other applications. Each row
has the three properties: the outer choice selects the inner alphabet,
filters come with the request, and a repeat is illegal. The first row is
this paper; a store without a parent chain is its own chain. Most rows
filter by location, and every row has at least one other filter. Each
filter is a sorted array of leaf ids, so filters combine by
intersection. Rows other than the first are illustrative; we did not
study them. *In development, not part of the benchmark.\par}
\vspace{0.8ex}
\footnotesize
\begin{adjustbox}{max width=\linewidth}
\begin{tabular}{@{}l l l l l l@{}}
\toprule
\thead{application} & \thead{outer slot} & \thead{inner slot} & \thead{location filter} & \thead{other filters} & \thead{emit once} \\
\midrule
food delivery & store & item & delivers to the cell & orderable* & chain
once, item once per row \\
e-commerce & seller & product & ships to the region & in stock & no
repeated product \\
travel & property & room type & near the destination & free on dates &
no repeated property \\
advertising & advertiser & creative & geographic targeting & budget left
& frequency cap per page \\
video feed & creator & video & licensed in region & not removed & no
repeated video \\
code completion & module & symbol & none & in import scope & no repeated
suggestion \\
database query & table & column & none & read permission & no repeated
column \\
\bottomrule
\end{tabular}
\end{adjustbox}
\end{table}

\subsection[I. Page length]{\hypertarget{sec:I}{}I. Page length}\label{i.-page-length}

With incremental state turned off, the decoder replays the prefix from
the root at every step. Sweeping page length then prices what carrying
the trie position forward is worth. This sweep is a side experiment,
outside the nine sweeps and the run counts of \hyperlink{tab:8}{Table 8}. The penalty grows
from nothing to nearly doubling the cost per token across an eightfold
range, and each doubling adds a roughly constant amount. The first point
is near zero, so the other three carry the trend; they cannot separate a
logarithm from a square root, but the per-token penalty grows more
slowly than page length.

\begin{table}[H]
\centering
{\small \hypertarget{tab:16}{}\textbf{Table 16.} Page length sweep, with incremental state turned off
so the prefix is replayed every step. The per-token penalty grows, but
the rate at which it grows falls by a factor of 3.4, so the per-token
penalty grows more slowly than page length. Token counts are this
sweep's own measurement, so twenty slots reads 809 here against the 843
token median of the main benchmark, and its +57.3 percent is the replay
ablation of Table 4 (+53.9) measured in this separate sweep.\par}
\vspace{0.8ex}
\footnotesize
\begin{adjustbox}{max width=\linewidth}
\begin{tabular}{@{}l r r l@{}}
\toprule
\thead{tokens per page} & \thead{store slots} & \thead{penalty from replaying the prefix} & \thead{rate over the previous interval} \\
\midrule
207 & 5 & −1.2\% & \\
409 & 10 & +28.3\% & 0.146 points per token \\
809 & 20 & +57.3\% & 0.073 points per token \\
1,608 & 40 & +91.8\% & 0.043 points per token \\
\bottomrule
\end{tabular}
\end{adjustbox}
\end{table}

\subsection[J. The implementations,
drawn]{\hypertarget{sec:J}{}J. The implementations,
drawn}\label{j.-the-implementations-drawn}

The body names each implementation in a sentence. These figures draw
what each one changes, on toys small enough to read in full. \hyperlink{fig:5}{Figures 5}
and \hyperlink{fig:6}{6} are the constraint and the decode step every implementation
shares; \hyperlink{fig:7}{Figures 7} to \hyperlink{fig:11}{11} are the changes the ranking compares.

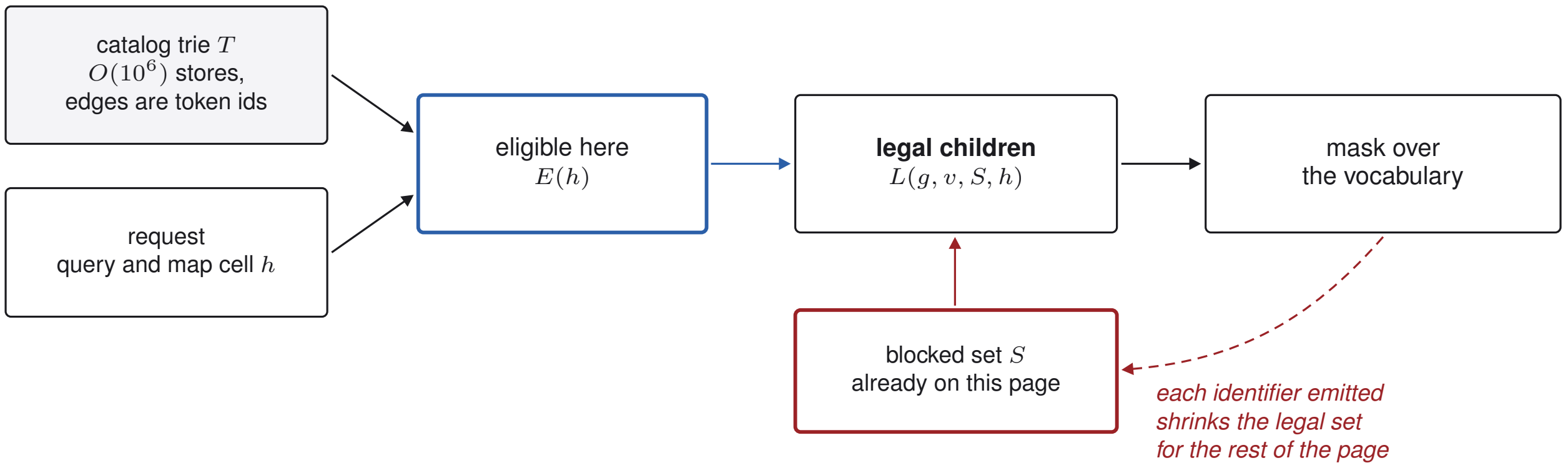
\begin{figure}[H]
\centering
\iflatexml\includegraphics[width=471.45pt]{figures/legalset.svg}\else
\begin{tikzpicture}[fig, x=0.925in, y=0.925in]
\FigBox[fill=figfaint]{0.06}{2.30}{1.45}{0.62}{catalog trie $T$\\$O(10^6)$ stores,\\edges are token ids}
\FigBox{0.06}{1.52}{1.45}{0.58}{request\\query and map cell $h$}
\FigBox[draw=figgpu, line width=1.1pt]{1.92}{1.90}{1.30}{0.62}{\figlabel eligible here\\\figlabel$E(h)$}
\FigBox{3.62}{1.90}{1.45}{0.62}{\figlabel\bfseries legal children\\\figlabel\bfseries$L(g, v, S, h)$}
\FigBox{5.47}{1.90}{1.60}{0.62}{\figlabel mask over\\\figlabel the vocabulary}
\FigBox[draw=figbad, line width=1.1pt]{3.62}{1.00}{1.45}{0.55}{blocked set $S$\\already on this page}

\draw[flow] (1.53, 2.61) -- (1.90, 2.35);
\draw[flow] (1.53, 1.81) -- (1.90, 2.07);
\draw[flow, figgpu] (3.24, 2.21) -- (3.60, 2.21);
\draw[flow] (5.09, 2.21) -- (5.45, 2.21);
\draw[flow, figbad, dash pattern=on 3pt off 2pt] (6.27, 1.88) to[bend left=22] (5.09, 1.28);
\draw[flow, figbad] (4.34, 1.57) -- (4.34, 1.88);
\FigNote[text=figbad, font=\figtext\itshape]{5.24}{1.04}{each identifier emitted\\shrinks the legal set\\for the rest of the page}

\foreach \i/\t in {0/{1. $L$ depends on $h$, so it differs per request. A per request compile would still be possible.},
                   1/{2. $L$ depends on $S$, so it differs per token inside one sequence. That rules the compile out.},
                   2/{3. $L$ is nested: the store sampled at the outer level decides which menu is used next.}}
  {\FigNote[text=figink, font=\figlabel]{0.06}{0.62-0.22*\i}{\t}}
\end{tikzpicture}
\fi
\par\vspace{0.8ex}
{\small \hypertarget{fig:5}{}\textbf{Figure 5. The legal set is assembled per request and then
narrowed per token.} The map cell selects the eligible store set through
the map-cell-to-store index. The blocked set then removes what the page
has already used. An ahead-of-time compile would need to fix both of
these, but only the first one is fixed when the request arrives.\par}
\end{figure}

\begin{figure}[H]
\centering
\iflatexml\includegraphics[width=477.38pt]{figures/axes.svg}\else
\begin{tikzpicture}[fig, x=0.925in, y=0.925in]
\def\H{3.68}
\def\GAP{0.26}  
\pgfmathsetmacro\TOP{\H-0.40}
\def\BH{0.72}

\foreach \x/\w/\f/\lab/\next in {
    0.05/1.38/white/{decode batch\\$n$ rows, each at its own\\position in its own page}/1,
    1.69/1.27/figfaint/{partition into $G$ groups\\by trie node, blocked set\\and grammar state}/1,
    3.22/1.19/white/{read the child list\\of each group's node}/1,
    4.67/1.19/white/{scatter the legal set\\into a mask buffer}/1,
    6.12/1.02/white/{apply the mask\\and sample}/0} {
  \FigBox[fill=\f]{\x}{\TOP-\BH}{\w}{\BH}{\lab}
  \ifnum\next=1 \draw[flow] ({\x+\w+0.02}, {\TOP-\BH/2}) -- ({\x+\w+\GAP-0.02}, {\TOP-\BH/2});\fi
}

\foreach \k/\name/\col/\xa/\xb in {0/placement/figgpu/3.22/5.86, 1/layout/figcpu/3.22/4.41,
                                   2/launch structure/figgood/4.67/7.14} {
  \pgfmathsetmacro\sy{\TOP-\BH-0.17-0.21*\k}
  \draw[\col, line width=1.8pt] (\xa, \sy) -- (\xb, \sy);
  \draw[\col, line width=1pt] (\xa, \sy-0.045) -- (\xa, \sy+0.045) (\xb, \sy-0.045) -- (\xb, \sy+0.045);
  \node[anchor=east, inner sep=0, font=\figlabel\bfseries, text=\col] at (\xa-0.10, \sy) {\name};
}
\FigNote[font=\figtext\itshape]{0.13}{\TOP-\BH-0.17-0.63+0.02}{Every axis acts on one or both of the two stages that run once per group, so a step pays for them $G$ times.}

\def\CY{1.38}
\node[inner sep=0, font=\large] at (3.6, \CY) {$C(\mathrm{step}) \;=\; G \cdot k \;+\; \sum_{\mathrm{groups}} O(d)$};
\node[inner sep=0, align=center, font=\figlabel] at (3.6, \CY-0.40) {$k$ is one kernel launch plus the CPU synchronization after it. $d$ is the mean child count.\\The first term is what the measurements are made of.};

\def\AY{0.10}
\def\AH{0.66}
\def\AW{2.26}
\foreach \i/\name/\col/\sub in {0/placement/figgpu/{does the walk execute on the GPU\\or on the CPU?},
                                1/layout/figcpu/{is the node structure padded to a\\fixed width, or ragged?},
                                2/launch structure/figgood/{is a launch paid once per group,\\or once per step?}} {
  \pgfmathsetmacro\x{0.09+\i*(\AW+0.11)}
  \FigBox[draw=\col, line width=1.1pt]{\x}{\AY}{\AW}{\AH}{}
  \node[inner sep=0, font=\figtitle, text=\col] at (\x+\AW/2, \AY+\AH-0.18) {\name};
  \node[inner sep=0, align=center] at (\x+\AW/2, \AY+0.22) {\sub};
}
\end{tikzpicture}
\fi
\par\vspace{0.8ex}
{\small \hypertarget{fig:6}{}\textbf{Figure 6. Where the three axes act on a decode step.} Placement
and layout change how the child list is read. Launch structure changes
how many times the mask work is handed to the GPU. Each acts on one or
both of the two stages that run once per group (the brackets mark
which), which is why none of them changes the arithmetic.\par}
\end{figure}

\begin{figure}[H]
\centering
\iflatexml\includegraphics[width=480.77pt]{figures/filters.svg}\else
\begin{tikzpicture}[fig, x=0.925in, y=0.925in]
\def\H{3.28}
\def\CW{0.27}
\def\CH{0.22}

\newcommand{\lane}[6]{%
  \pgfmathsetmacro\Y{#1}%
  \node[anchor=west, inner sep=0, font=\figlabel\bfseries, text=#2] at (0.05, \Y+0.02) {#3};
  \FigBox{0.05}{\Y-0.72}{1.20}{0.46}{#4}
  \draw[flow] (1.27, \Y-0.49) -- (1.55, \Y-0.49);
  \FigBox{1.57}{\Y-0.72}{1.35}{0.46}{#5}
  \draw[flow] (2.94, \Y-0.49) -- (3.22, \Y-0.37);
  \FigNote[text=#2]{3.25}{\Y-0.14}{eligible E}
  \FigNote[text=figbad]{3.25}{\Y-0.62}{#6}
  \FigBox{5.12}{\Y-0.86}{2.08}{0.62}{legal child test, in each child's [lo, hi):\\count of E\enspace$>$\enspace count of S $\cap$ E\\each count: two binary searches}}
\newcommand{\cellrow}[3]{%
  \def\n{0}%
  \foreach \v [count=\c from 0] in {#2} {\FigCell[#3]{3.25+\c*\CW}{#1}{\CW}{\CH}{\v}\xdef\n{\c}}%
  \pgfmathsetmacro\E{3.25+(\n+1)*\CW}}

\node[font=\figtitle, anchor=west, inner sep=0] at (0.05, \H-0.12) {Every filter becomes a sorted array of leaf ids, and the test counts in two of them};

\lane{\H-0.42}{figslate}{Outer slot: which store}{request's\\map cell h}{map cell $\to$ store index\\(sorted store ids\\per cell)}{blocked S: filled stores,\\each with its chain}
\cellrow{\Y-0.37}{3, 8, 9, 14, 21, 40}{}
\draw[flow] (\E+0.05, \Y-0.37) -- (5.10, \Y-0.52);
\cellrow{\Y-0.93}{8, 9}{draw=figbad, fill=figtint}
\draw[flow] (\E+0.05, \Y-0.93) -- (5.10, \Y-0.62);

\draw[flow] (6.16, \H-1.30) -- (6.16, \H-1.62);
\node[anchor=east, inner sep=0] at (6.08, \H-1.46) {the walk ends on a store leaf, say 21};

\lane{\H-1.80}{figpurple}{Inner slot: which item}{store leaf 21,\\just chosen}{store to item index\\(sorted item ids per store)}{blocked S: items already\\in this row}
\cellrow{\Y-0.37}{5, 17, 30, 33}{}
\draw[flow] (\E+0.05, \Y-0.37) -- (5.10, \Y-0.52);
\cellrow{\Y-0.93}{17}{draw=figbad, fill=figtint}
\draw[flow] (\E+0.05, \Y-0.93) -- (5.10, \Y-0.62);

\FigNote{0.05}{0.14}{Filters combine by intersecting sorted arrays, so a new filter is one more array, not a new automaton. Chain exclusion enters through S:\\a filled store adds every store of its chain. Store 8 was placed, and 9 shares its chain.}
\end{tikzpicture}
\fi
\par\vspace{0.8ex}
{\small \hypertarget{fig:7}{}\textbf{Figure 7. How the request's filters reach the legal child test.}
Each filter is a sorted array of leaf ids. For an outer slot the
eligible array comes from the map-cell-to-store index and the blocked
array from the stores already on the page, each widened to its chain;
for an inner slot the eligible array is the chosen store's menu and the
blocked array the items already in its carousel. The test is the one of
Figure 3.\par}
\end{figure}

\begin{figure}[H]
\centering
\iflatexml\includegraphics[width=447.11pt]{figures/expansion.svg}\else
\begin{tikzpicture}[fig, x=0.925in, y=0.925in]
\def\H{4.10}
\def\CW{0.27}  
\def\CH{0.22}  
\newcommand{\mono}{\fontsize{7}{8}\selectfont\ttfamily}

\node[font=\figtitle, anchor=west, inner sep=0] at (0.05, \H-0.12) {Three rows of a batch, at three nodes of the five leaf trie};
\foreach \x/\head in {0.10/row, 0.75/at node, 1.55/{its edges, one offsets slice}, 3.25/tokens}
  {\FigNote{\x}{\H-0.40}{\head}}
\foreach \r/\at/\edges/\toks in {0/root/{0, 1}/{a, b}, 1/node 1/{2, 3, 4}/{x, y, z}, 2/node 2/{5, 6}/{x, y}} {
  \pgfmathsetmacro\y{\H-0.60-0.18*\r}
  \FigNote[text=figink]{0.10}{\y}{row \r}
  \FigNote[text=figink, font=\mono]{0.75}{\y}{\at}
  \FigNote[text=figink, font=\mono]{1.55}{\y}{\edges}
  \FigNote[text=figink, font=\mono]{3.25}{\y}{\toks}
}

\pgfmathsetmacro\YA{\H-1.30}
\node[font=\figtitle, anchor=west, inner sep=0] at (0.05, \YA) {A\enspace Padded: every row expanded to max\_children};
\foreach \r/\start/\width in {0/0/2, 1/2/3, 2/5/2} {
  \pgfmathsetmacro\y{\YA-0.34-0.25*\r}
  \node[anchor=east, inner sep=0, font=\mono] at (0.57, \y) {row \r};
  \foreach \c in {0,...,7} {
    \pgfmathtruncatemacro\v{\start+\c}
    \ifnum\c<\width \FigCell{0.65+\c*\CW}{\y}{\CW}{\CH}{\v}%
    \else \FigCell[padcell]{0.65+\c*\CW}{\y}{\CW}{\CH}{}\fi
  }
  \node[inner sep=0, text=fignil] at (0.65+8*\CW+0.10, \y) {\dots};
  \FigCell[padcell]{0.65+8*\CW+0.25}{\y}{0.30}{\CH}{}
}
\FigNote{0.65}{\YA-1.22}{one fixed shape, \ValPaddedSlots\ slots wide on the real tries;\\a valid mask marks the real children (white).\\The shape is known in advance, so no CPU read.}

\def\XB{4.00}
\node[font=\figtitle, anchor=west, inner sep=0] at (\XB, \YA) {B\enspace Ragged: only the real children, in one flat list};
\node[anchor=east, inner sep=0, font=\mono] at (\XB+0.67, \YA-0.34) {edge};
\node[anchor=east, inner sep=0, font=\mono] at (\XB+0.67, \YA-0.59) {row};
\foreach \c/\row in {0/0, 1/0, 2/1, 3/1, 4/1, 5/2, 6/2} {
  \FigCell{\XB+0.75+\c*\CW}{\YA-0.34}{\CW}{\CH}{\c}
  \FigCell[draw=figcpu, fill=figcputint]{\XB+0.75+\c*\CW}{\YA-0.59}{\CW}{\CH}{\row}
}
\FigNote{\XB}{\YA-1.02}{Its length, 2 + 3 + 2 = 7, must be known to allocate it,\\so the CPU reads it. Free on the CPU; on the GPU\\a read back, once per group per step.}

\def\TY{1.10}
\node[font=\figtitle, anchor=west, inner sep=0] at (0.05, \TY) {C\enspace What each combination pays, light load};
\FigNote[text=figink]{1.25}{\TY-0.24}{padded}
\FigNote[text=figink]{4.20}{\TY-0.24}{ragged}
\FigNote[text=figink]{0.10}{\TY-0.52}{on the GPU}
\FigNote[text=figgpu]{1.25}{\TY-0.52}{no read back; padding is cheap for\\lockstep threads. The control, 0\%}
\FigNote[text=figbad]{4.20}{\TY-0.52}{a read back per group per step:\\\ValRaggedGpu\%}
\FigNote[text=figink]{0.10}{\TY-0.88}{on the CPU}
\FigNote[text=figbad]{1.25}{\TY-0.88}{walks mostly padding: about \ValPaddedWalkSlowdown\ times slower\\in an isolated walk, \ValPaddedCpu\% end to end}
\FigNote[text=figgood]{4.20}{\TY-0.88}{walks only real children, no read back:\\the CPU trie, \ValRaggedCpu\%}
\end{tikzpicture}
\fi
\par\vspace{0.8ex}
{\small \hypertarget{fig:8}{}\textbf{Figure 8. Padded or ragged child expansion.} Both layouts store
the same compressed sparse row arrays; they differ in how a group's
child lists are expanded when its mask is built. (A) Padded gives every
row the width of the widest node, a fixed 8,192 slots on the real tries,
so the shape is known in advance and the GPU needs no read-back. (B)
Ragged lists only the real children, but its length must be read on the
CPU before the list can be allocated, which on the GPU is a read-back
once per group per step. (C) Each layout suits one placement, which is
the interaction of Table 3.\par}
\end{figure}

\begin{figure}[H]
\centering
\iflatexml\includegraphics[width=482.82pt]{figures/compaction.svg}\else
\begin{tikzpicture}[fig, x=0.925in, y=0.925in]
\def\H{3.55}
\pgfmathsetmacro\MY{\H-1.30}  
\pgfmathsetmacro\LY{\H-2.05}  
\pgfmathsetmacro\RY{\H-0.55}  

\newcommand{\tnode}[5][]{%
  \path[draw=figink, fill=white, line width=0.8pt, #1] ({#2}, {#3}) circle ({#4});
  \node[inner sep=0, font=\figsmall] at ({#2}, {#3}) {#5};}
\newcommand{\tedge}[7]{%
  \draw[figink, line width=0.6pt] ({#1}, {#2-#5}) -- ({#3}, {#4+#6})
    node[pos=0.58, fill=white, inner sep=0.8pt, font=\figlabel\itshape, text=figpurple] {#7};}

\node[font=\figtitle, anchor=west, inner sep=0] at (0.05, \H-0.12) {A\enspace The item trie, with one store's menu marked};
\tnode{1.75}{\RY}{0.17}{[0,8)}
\foreach \x/\tok/\lab in {0.75/a/{[0,3)}, 1.75/b/{[3,5)}, 2.85/c/{[5,8)}} {
  \tedge{1.75}{\RY}{\x}{\MY}{0.17}{0.17}{\tok}
  \tnode{\x}{\MY}{0.17}{\lab}
}
\foreach \i/\x/\px/\tok/\on in {0/0.25/0.75/x/0, 1/0.60/0.75/y/1, 2/0.95/0.75/z/1, 3/1.55/1.75/x/0,
                                4/1.95/1.75/y/0, 5/2.45/2.85/x/1, 6/2.85/2.85/y/0, 7/3.25/2.85/z/0} {
  \tedge{\px}{\MY}{\x}{\LY}{0.17}{0.13}{\tok}
  \ifnum\on=1 \tnode[draw=figpurple, line width=1.3pt]{\x}{\LY}{0.13}{}%
  \else \tnode[draw=fignil, line width=0.6pt]{\x}{\LY}{0.13}{}\fi
  \node[inner sep=0, font=\figsmall] at (\x, \LY-0.25) {\i};
}
\FigNote[text=figpurple]{0.05}{\LY-0.50}{purple: on the menu of the chosen store, leaves 1, 2 and 5}
\FigNote[text=figink]{0.05}{\LY-0.72}{Legal test per child edge, every step:}
\FigNote[text=figink, font=\fontsize{7}{8}\selectfont\ttfamily]{0.05}{\LY-0.92}{on menu in [lo, hi) = search(menu, hi) $-$ search(menu, lo)}
\FigNote{0.05}{\LY-1.12}{two binary searches, repeated at every step of the row}

\draw[thickflow] (3.55, \MY) -- (4.15, \MY);
\FigNote{3.52}{\MY-0.36}{built once when the\\store's row opens,\\cached per menu}

\def\XC{4.35}
\node[font=\figtitle, anchor=west, inner sep=0] at (\XC, \H-0.12) {B\enspace The compacted trie: only the menu, renumbered};
\tnode{5.65}{\RY}{0.17}{[0,3)}
\tedge{5.65}{\RY}{5.05}{\MY}{0.17}{0.17}{a}
\tnode{5.05}{\MY}{0.17}{[0,2)}
\tedge{5.65}{\RY}{6.35}{\MY}{0.17}{0.17}{c}
\tnode{6.35}{\MY}{0.17}{[2,3)}
\foreach \i/\x/\px/\tok/\old in {0/4.75/5.05/y/1, 1/5.35/5.05/z/2, 2/6.35/6.35/x/5} {
  \tedge{\px}{\MY}{\x}{\LY}{0.17}{0.13}{\tok}
  \tnode[draw=figpurple, line width=1.3pt]{\x}{\LY}{0.13}{}
  \node[inner sep=0, font=\figsmall] at (\x, \LY-0.25) {\i\enspace(was \old)};
}
\FigNote[text=figink]{\XC}{\LY-0.50}{every leaf is on the menu, so the same count is}
\FigNote[text=figink, font=\fontsize{7}{8}\selectfont\ttfamily]{\XC}{\LY-0.72}{on menu in [lo, hi) = hi $-$ lo}
\FigNote{\XC}{\LY-0.98}{one subtraction. Blocked items are translated to the new\\ids once; an item off the menu is dropped.}
\end{tikzpicture}
\fi
\par\vspace{0.8ex}
{\small \hypertarget{fig:9}{}\textbf{Figure 9. Inner subtrie compaction.} When a store's carousel
opens, the decoder builds a trie over only that store's menu and
renumbers its leaves from zero, then walks the small trie for the rest
of the carousel. Every leaf of the small trie is on the menu, so
counting eligible leaves in a node's range becomes a subtraction instead
of two binary searches per child edge. Outer subtrie compaction does the
same over the stores of the request's map cell; it had no significant
effect on the GPU, and on the CPU it never ran in the same round as its
base, so we cannot separate it from the base. With it, the CPU trie is
16.7 percent faster than the control at light load and 55.1 percent
faster when saturated (Table 4).\par}
\end{figure}

\begin{figure}[H]
\centering
\iflatexml\includegraphics[width=439.17pt]{figures/incremental.svg}\else
\begin{tikzpicture}[fig, x=0.925in, y=0.925in]
\def\H{3.55}
\def\XO{1.45}  
\def\CW{0.32}  

\node[font=\figtitle, anchor=west, inner sep=0] at (0.05, \H-0.12) {A\enspace Blocked set memo: rebuild the blocked set only when a slot closes};
\FigNote{0.05}{\H-0.42}{decode step}
\foreach \t in {1,...,15}
  \node[inner sep=0, font=\figsmall, text=figmuted] at ({\XO+(\t-0.5)*\CW}, \H-0.42) {\t};
\foreach \t in {5,10,15}
  \draw[figbad, line width=0.6pt, dash pattern=on 2.5pt off 1.5pt] ({\XO+\t*\CW}, \H-1.26) -- ({\XO+\t*\CW}, \H-0.55);
\FigNote[text=figink]{0.05}{\H-0.72}{without memo}
\FigNote[text=figink]{0.05}{\H-1.08}{with memo}
\foreach \t in {1,...,15} {
  \pgfmathsetmacro\x{\XO+(\t-1)*\CW+0.02}
  \path[fill=figcpu, draw=figcpu, line width=0.5pt] (\x, \H-0.83) rectangle ++(\CW-0.04, 0.22);
  \pgfmathtruncatemacro\closes{mod(\t,5)==0}
  \ifnum\closes=1
    \path[fill=figcpu, draw=figcpu, line width=0.5pt] (\x, \H-1.19) rectangle ++(\CW-0.04, 0.22);
  \else
    \path[fill=white, draw=figgrid, line width=0.5pt] (\x, \H-1.19) rectangle ++(\CW-0.04, 0.22);
  \fi
}
\FigNote{0.05}{\H-1.52}{Dashed: a slot closes (every identifier is five tokens). Orange: expand every filled\\slot to its chain and sort. The memo keeps each row's set and adds only the new slot.}

\node[font=\figtitle, anchor=west, inner sep=0] at (0.05, \H-1.98) {B\enspace Incremental state: keep the trie position between steps};
\pgfmathsetmacro\ya{\H-2.38}
\pgfmathsetmacro\yb{\H-2.84}
\FigNote[text=figink]{0.05}{\ya}{replay the prefix}
\FigNote[text=figink]{0.05}{\yb}{incremental}
\foreach \i/\tok in {0/a, 1/x, 2/b, 3/y, 4/\dots} {
  \pgfmathsetmacro\xa{1.60+0.66*\i}
  \draw[figbad, line width=0.7pt, -{Triangle[length=3.5pt, width=3.2pt]}] (\xa, \ya) -- (\xa+0.64, \ya);
  \node[inner sep=0, font=\figlabel\itshape, text=figgpu] at (\xa+0.33, \ya+0.11) {\tok};
}
\fill[figink] (1.60, \ya) circle (1.6pt);
\node[inner sep=0, font=\figsmall, text=figmuted] at (1.60, \ya-0.15) {root};
\draw[figgood, line width=0.7pt, -{Triangle[length=3.5pt, width=3.2pt]}] (1.60+4*0.66, \yb) -- (1.60+5*0.66-0.02, \yb);
\fill[figink] (1.60+4*0.66, \yb) circle (1.6pt);
\node[inner sep=0, font=\figsmall, text=figmuted] at (1.60+4*0.66, \yb-0.15) {saved position};
\FigNote{5.25}{\ya}{every step re-walks the page\\so far: work grows with\\page length}
\FigNote{5.25}{\yb}{every step advances\\one edge}

\FigNote{0.05}{\H-3.30}{Replaying costs \ValReplayLight\ percent at light load and \ValReplaySaturated\ percent when saturated,\\and \ValReplayTwentySlots\ percent at twenty slots in the page length sweep.}
\end{tikzpicture}
\fi
\par\vspace{0.8ex}
{\small \hypertarget{fig:10}{}\textbf{Figure 10. Two things the decoder does not recompute every
token.} (A) The blocked set changes only when a slot closes, so the memo
keeps it per row and adds each new slot, instead of expanding every
filled slot to its chain and sorting on every step. (B) Each row keeps
its position in the trie, so a step advances one edge; the ablation
replays the whole page prefix every step, and its cost grows with page
length (\hyperlink{tab:16}{Table 16}).\par}
\end{figure}

\begin{figure}[H]
\centering
\iflatexml\includegraphics[width=477.76pt]{figures/cpustack.svg}\else
\begin{tikzpicture}[fig, x=0.925in, y=0.925in]
\def\H{3.95}
\def\XO{1.45}  
\def\U{0.62}   

\node[font=\figtitle, anchor=west, inner sep=0] at (0.05, \H-0.12) {A\enspace Overlap: walk every group while the model runs its forward pass};
\foreach \k/\lab/\start in {0/CPU recipe/3.0, 1/+ overlap/0.25} {
  \pgfmathsetmacro\yg{\H-0.48-0.62*\k}
  \pgfmathsetmacro\yh{\H-0.70-0.62*\k}
  \FigNote[text=figink]{0.05}{(\yg+\yh)/2}{\lab}
  \node[anchor=east, inner sep=0, font=\figsmall, text=figmuted] at (\XO-0.08, \yg) {GPU};
  \node[anchor=east, inner sep=0, font=\figsmall, text=figmuted] at (\XO-0.08, \yh) {CPU};
  \FigBar[fill=fignil, draw=fignil]{\XO}{\XO+3*\U}{\yg}{0.18}{\color{white}model forward}
  \ifnum\k=0
    \FigBar[fill=white, draw=figgrid]{\XO}{\XO+3*\U}{\yh}{0.18}{\color{figmuted}blocked: copy queued behind the forward}
  \fi
  \foreach \i in {0,1,2} {
    \pgfmathsetmacro\a{\XO+(\start+0.5*\i)*\U}
    \FigBar[fill=figcpu, draw=figcpu]{\a+0.02}{\a+0.5*\U-0.02}{\yh}{0.18}{\color{white}g\the\numexpr\i+1\relax}
  }
  \pgfmathsetmacro\ap{max(3.0, \start+1.5)}
  \FigBar[fill=figgpu, draw=figgpu]{\XO+\ap*\U+0.03}{\XO+(\ap+0.55)*\U}{\yg}{0.18}{\color{white}apply}
  \draw[figbad, line width=0.7pt, dash pattern=on 2.5pt off 1.5pt]
    ({\XO+(\ap+0.55)*\U+0.04}, {\yh-0.12}) -- ({\XO+(\ap+0.55)*\U+0.04}, {\yg+0.12});
}
\FigNote{\XO+5.45*\U}{\H-0.59}{the step ends when the\\last walk ends}
\FigNote{\XO+5.45*\U}{\H-1.21}{the walks hide under the\\forward; masks go through\\pinned memory, no blocking}

\pgfmathsetmacro\yB{\H-1.75}
\node[font=\figtitle, anchor=west, inner sep=0] at (0.05, \yB) {B\enspace Numpy trie: the same steps as numpy calls, from a cached state};
\pgfmathsetmacro\ya{\yB-0.40}
\pgfmathsetmacro\yb{\yB-0.80}
\FigNote[text=figink]{0.05}{\ya}{tensor trie\\{\figsmall\color{figbad}torch calls, $\approx$\,10\,\textmu s each, est.}}
\FigNote[text=figink]{0.05}{\yb}{numpy trie\\{\figsmall\color{figgood}numpy calls, $\approx$\,1\,\textmu s each, est.}}
\def\xs{1.55}\def\st{0.90}\def\bw{0.80}\def\bh{0.25}
\foreach \i/\lab/\sty in {0/walk from root/tintbox, 1/expand children/tintbox, 2/{gather lo, hi}/gpubox,
                          3/{count, compare}/gpubox, 4/scatter to mask/tintbox} {
  \FigBox[\sty]{\xs+\i*\st}{\ya-\bh/2}{\bw}{\bh}{\figsmall\lab}
}
\foreach \i/\lab/\sty in {0/cached state/cputintbox, 1/slice children/cputintbox, 2/{gather lo, hi}/cpubox,
                          3/{count, compare}/cpubox, 4/filter to list/cputintbox} {
  \FigBox[\sty]{\xs+\i*\st}{\yb-\bh/2}{\bw}{\bh}{\figsmall\lab}
}
\foreach \y in {\ya, \yb} {
  \foreach \i in {1,...,5} {
    \draw[thickflow] ({\xs+(\i-1)*\st+\bw+0.01}, \y) -- ({\xs+\i*\st-0.01-(\i==5)*0.02}, \y);
  }
}
\pgfmathsetmacro\xe{\xs+5*\st}
\FigBox[tintbox]{\xe+0.06}{\ya-\bh/2}{1.05}{\bh}{\figsmall vocabulary wide mask}
\foreach \c/\v in {0/17, 1/30, 2/33} {
  \FigCell{\xe+0.21+\c*0.30}{\yb}{0.30}{0.22}{\v}
}

\pgfmathsetmacro\yC{\yB-1.25}
\node[font=\figtitle, anchor=west, inner sep=0] at (0.05, \yC) {C\enspace Batched write: the step's masks leave in one copy, not one per group};
\pgfmathsetmacro\ya{\yC-0.36}
\pgfmathsetmacro\yb{\yC-0.72}
\FigNote[text=figink]{0.05}{\ya}{per group}
\FigNote[text=figink]{0.05}{\yb}{batched}
\foreach \g in {0,1,2} {
  \pgfmathsetmacro\x{1.45+\g*1.06}
  \FigBar[fill=figcpu, draw=figcpu]{\x}{\x+0.30}{\ya}{0.18}{\color{white}copy}
  \foreach \k in {0,1,2} {
    \FigBar[fill=figgpu, draw=figgpu]{\x+0.32+\k*0.18}{\x+0.48+\k*0.18}{\ya}{0.18}{}
  }
}
\FigNote{1.45+3*1.06+0.05}{\ya}{a copy and several kernels\\for every group}
\FigBar[fill=figcpu, draw=figcpu]{1.45}{2.35}{\yb}{0.18}{\color{white}one pinned copy}
\foreach \k in {0,1,2} {
  \FigBar[fill=figgpu, draw=figgpu]{2.39+\k*0.18}{2.55+\k*0.18}{\yb}{0.18}{}
}
\FigNote{2.39+3*0.18+0.10}{\yb}{a fixed set of kernels per step,\\however many groups}
\FigBar[fill=figcpu, draw=figcpu]{6.10}{6.25}{\ya+0.02}{0.12}{}
\FigNote[text=figink, font=\figsmall]{6.29}{\ya+0.02}{CPU to GPU copy}
\FigBar[fill=figgpu, draw=figgpu]{6.10}{6.25}{\ya-0.18}{0.12}{}
\FigNote[text=figink, font=\figsmall]{6.29}{\ya-0.18}{GPU kernel}
\end{tikzpicture}
\fi
\par\vspace{0.8ex}
{\small \hypertarget{fig:11}{}\textbf{Figure 11. The full CPU stack: three changes on top of the CPU
recipe.} Schematic. (A) The overlap walks every group before the first
copy, so the walks run under the forward pass and the step no longer
waits for them. (B) The tensor trie re-walks the current identifier's
few prefix tokens from the trie root on every query; the page prefix is
incremental in both, so this is not the ablation of \hyperlink{fig:10}{Figure 10}(B). The
numpy trie runs the same steps from a cached state, slicing the child
edges instead of expanding them, as numpy calls at about 1 microsecond
each instead of about 10 by our estimate of dispatch overhead, and
returns a short list of allowed ids instead of a vocabulary wide mask.
Filled boxes mark the steps that differ. (C) The batched write sends
every group's rows in one pinned copy and a fixed set of kernels per
step. None changes which tokens are allowed; together they take the CPU
recipe's 20.1 percent at light load to 65.9 percent (Tables 4, \hyperlink{tab:9}{9} and
\hyperlink{tab:17}{17}).\par}
\end{figure}

\subsection[K. The full CPU stack's
options]{\hypertarget{sec:K}{}K. The full CPU stack's
options}\label{k.-the-full-cpu-stacks-options}

Figure 4(c) shows a wait the CPU recipe does not need: each group opens
with a blocking copy of its row indices, which queues behind the
forward, so the first walk starts only when the forward ends. The full
CPU stack adds three options to the CPU recipe, each on top of the
previous one, and none changes which tokens are allowed. The overlap
walks every group before the first copy and sends the masks through
pinned memory with copies that do not block, so the walks run during the
forward (Figure 4(d)). The numpy trie answers the same trie queries over
the same compressed sparse row arrays at about 1 microsecond each
instead of about 10, an estimate of the tensor calls' dispatch overhead
rather than a measurement. It caches the state each identifier prefix
reached, where the CPU recipe's tensor trie (torch operations on the
same arrays) re-walks those few tokens from the root, and returns short
lists of allowed tokens instead of vocabulary-wide masks. The batched
write sends every group's rows in one pinned copy and writes them with a
fixed set of kernels per step.

The options ran in sweeps 8 and 9, on a newer build than sweeps 1 to 7,
with twelve rounds per regime and a 4 billion parameter model beside the
0.6 billion one. Each arm is compared with the control of its own round.
The CPU recipe ran on both builds and saves 18.7 and 62.6 percent in
sweep 7 against 20.1 and 63.3 in sweep 8, so the newer build does not
move it. Sweep 9 measured the numpy trie and the batched write on top of
the overlap, on the 0.6 billion parameter model, twelve rounds per
regime, with no control. We pair the full CPU stack with sweep 9's
overlap arm in the same round and compose it with the overlap's effect,
multiplying the two ratios and combining their intervals on the log
scale. On the item-first layout we compare it with that layout's earlier
control instead, by a ratio of geometric means over rounds with a Welch
interval. Neither interval covers the difference between builds, so we
read these numbers as indicative. Composed, it is 65.9 percent faster
than the control at light load and 97.9 percent faster when saturated,
where it serves 96.9 percent more pages per second than sweep 8's
control (0.350), about 0.69 against the 0.7 arrival rate. Once an
implementation finishes pages faster than they arrive, the saturated
regime measures the queue it removed rather than the decoder, and the
light regime is the measure to read. At light load the constraint's own
cost falls from about 8.2 to about 1.4 ms per token, a budget that takes
the model's 2.1 ms from sweeps 1 to 7.

\begin{table}[H]
\centering
{\small \hypertarget{tab:17}{}\textbf{Table 17.} The full CPU stack's options, time per token. Each
cell is percent change against the control, with a 95 percent interval
over twelve rounds; the control column is absolute. The last column
compares the overlap with the recipe directly. These runs used a newer
build than sweeps 1 to 7, so each arm is paired only with the control of
its own round. Lower is faster.\par}
\vspace{0.8ex}
\footnotesize
\begin{adjustbox}{max width=\linewidth}
\begin{tabular}{@{}l l r r r r@{}}
\toprule
\thead{model} & \thead{regime} & \thead{control, ms per token} & \thead{CPU recipe} & \thead{recipe + overlap} & \thead{overlap against the recipe} \\
\midrule
0.6B & light & 10.30 {[}10.05, 10.54{]} & −20.1\% {[}−22.1, −18.1{]} &
−34.5\% {[}−37.0, −32.1{]} & −18.1\% {[}−20.3, −15.9{]} \\
0.6B & saturated & 187.98 {[}178.58, 197.38{]} & −63.3\% {[}−67.8,
−58.8{]} & −77.6\% {[}−80.7, −74.4{]} & −37.3\% {[}−48.3, −26.4{]} \\
4B & light & 13.80 {[}13.36, 14.24{]} & −28.9\% {[}−31.4, −26.4{]} &
−48.5\% {[}−50.5, −46.5{]} & −27.5\% {[}−28.9, −26.2{]} \\
4B & saturated & 222.80 {[}214.76, 230.84{]} & −63.6\% {[}−66.1,
−61.0{]} & −76.5\% {[}−78.5, −74.4{]} & −35.0\% {[}−40.9, −29.0{]} \\
\bottomrule
\end{tabular}
\end{adjustbox}
\end{table}

\subsection[L. The concurrency ladder]{\hypertarget{sec:L}{}L. The concurrency ladder}\label{l.-the-concurrency-ladder}

We reran the CPU trie against the control with a cap on how many rows
one decode step may hold. A first version of this ladder seemed to climb
from 5.4 percent at a batch cap of 1 to 32.8 percent at 256. That climb
is an artifact: some rungs ran before the cap was forwarded to the
serving engine, so they executed at the default batch size whatever rung
they were filed under. With those runs removed, the CPU trie saves
nothing at a cap of 1, 7.7 percent at 4, and 24 to 25 percent at 16, 64
and 256. At a cap of 1 there is one group, so there is no per group cost
to remove. A flat curve fits a fixed cost shared across the batch, and
fits GPU contention badly. Prefetching the mask on a side stream, which
should help if contention were the cause, is a 14.1 percent loss at
light load in the ranking. These figures come from the ladder run itself
and are not regenerated on each rebuild.

\subsection[M. Grammar engine
encodings]{\hypertarget{sec:M}{}M. Grammar engine
encodings}\label{m.-grammar-engine-encodings}

XGrammar precomputes masks for tokens whose validity does not depend on
context and computes only the rest at runtime. A LEGIT legal set depends
on the requester's map cell, so it is not known when the grammar
compiles, and on what the page has already emitted, so it is not stable
within one request. We tried the three available encodings (\hyperlink{tab:18}{Table 18}).

\begin{table}[H]
\centering
{\small \hypertarget{tab:18}{}\textbf{Table 18.} Three encodings of this constraint for an ahead of
time grammar engine. Exact, compilable, fast: any two.\par}
\vspace{0.8ex}
\footnotesize
\begin{adjustbox}{max width=\linewidth}
\begin{tabular}{@{}l l l l@{}}
\toprule
\thead{encoding} & \thead{constraint correct} & \thead{compiles} & \thead{fast} \\
\midrule
exact grammar, one rule per trie node & yes & no & untested \\
factored grammar, one rule per level & no & yes & yes \\
interpreter loop in Python on the CPU & yes & yes & no, about 100 times
slower \\
\bottomrule
\end{tabular}
\end{adjustbox}
\end{table}

\begin{itemize}
\tightlist
\item
  \textbf{The exact grammar} has O(10⁷) rules and hundreds of megabytes.
  Compiling it exhausts memory and time.
\item
  \textbf{The factored grammar} accepts the full cross product, about 6
  × 10¹⁷ item tuples, so greedy decoding lands off catalog and validity
  collapses.
\item
  \textbf{The interpreter loop} runs about 105 tokens per second per
  actor and uses only about 0.4 percent of the GPU. Its key value cache
  had 655 times the headroom it needed, so it was waiting on the CPU,
  not on batch size or memory.
\end{itemize}

\end{document}